\documentclass[twocolumn,prl,amsmath,amssymb,superscriptaddress]{revtex4-1}
\usepackage{graphicx}
\usepackage{dcolumn}
\usepackage{bm}
\usepackage{mathrsfs}
\usepackage{color}
\usepackage{epstopdf}
\usepackage{booktabs}
\usepackage{amsmath}
\usepackage[T1]{fontenc}
\usepackage{calligra}
\usepackage{anyfontsize}
\usepackage[normalem]{ulem}

\usepackage{physics}

\newcommand{\innerp}[2]{\left\langle #1 \vert #2 \right\rangle}

\newcommand{\be}{\begin{equation}}
\newcommand{\ee}{\end{equation}}
\newcommand{\bea}{\begin{eqnarray}}
\newcommand{\eea}{\end{eqnarray}}

\newcommand{\A}{{\rm A}}
\newcommand{\B}{{\rm B}}

\begin{document}

\title{Dual-wavelength quantum skyrmions from liquid crystal topological defects}

\author{Mwezi Koni}
\thanks{These authors contributed equally to this work.}
 \affiliation{School of Physics, University of the Witwatersrand, Private Bag 3, Wits 2050, South Africa}
 
\author{Fazilah Nothlawala}
\thanks{These authors contributed equally to this work.}
\affiliation{School of Physics, University of the Witwatersrand, Private Bag 3, Wits 2050, South Africa}

\author{Vagharshak Hakobyan}
\affiliation{University of Bordeaux, CNRS, Laboratoire Ondes et Mati{\`e}re d'Aquitaine, Talence, France}

\author{Isaac Nape}
\affiliation{School of Physics, University of the Witwatersrand, Private Bag 3, Wits 2050, South Africa}

\author{Etienne Brasselet}
\email{etienne.brasselet@u-bordeaux.fr}
\affiliation{University of Bordeaux, CNRS, Laboratoire Ondes et Mati{\`e}re d'Aquitaine, Talence, France}

\author{Andrew Forbes}
\email{andrew.forbes@wits.ac.za}
\affiliation{School of Physics, University of the Witwatersrand, Private Bag 3, Wits 2050, South Africa}

\date{\today}

\begin{abstract}
\noindent We propose a spin-orbit strategy for generating dual-wavelength quantum skyrmions, realized either as entangled photon pairs at dual wavelengths or as heralded single-photon states at a given wavelength---regimes neither previously conceptualized nor demonstrated. By coupling a two-photon entangled state to an electrically tunable liquid crystal topological defect, we engineer both nonlocal and local skyrmionic topologies in a reconfigurable platform. We highlight with examples how this approach may open new directions for engineered topological quantum states that exploit the topological richness of liquid crystals.
\end{abstract}

\maketitle

\noindent Skyrmions are particle-like topological objects, defined as maps between spheres and characterized by an integer Skyrme number that counts how many times the target sphere is wrapped. They have been realized in various physical systems, including atomic matter \cite{leslie2009creation}, chiral liquid crystals \cite{smalyukh2010three,chen2013generating}, acoustics \cite{ge2021observation, muelas2022observation}, and water waves \cite{wang2025topological}. In classical optics, skyrmions are formed by combining polarization and orbital angular momentum (OAM) into spin-textured fields, and have been demonstrated in evanescent waves \cite{tsesses2018optical}, paraxial beams \cite{gutierrez2021optical, gao2020paraxial}, and non-paraxial light \cite{du2019deep, sugic2021particle}. Their interaction with matter reveals rich physical phenomena \cite{kuratsuji2021evolution, chen2025more}, and enables robustness to noise and complex environments \cite{wang2024topological, ornelas2025topological}. Optical skyrmions have been generated using metasurfaces \cite{shen2024topologically}, graded-index lenses \cite{he2024optical}, on-chip devices \cite{lin2024chip}, and spatial light modulators \cite{shen2022generation}, with recent advances demonstrating topological transfer between light and matter \cite{mitra2025topological, lin2024wavelength}.

Quantum optical skyrmions were recently demonstrated both at the single-photon level \cite{liu2025nanophotonic} and in entangled photon pairs \cite{ornelas2024non}, opening new avenues for topological quantum optics. However, these realizations remain limited to single-wavelength configurations, with no ability to switch between entangled and single-photon topological textures  within a single platform, and do not extend to heralded single-photon states. Here, we report optical skyrmions in both entangled photon pairs and heralded single photons at distinct wavelengths, using a single reconfigurable device. We demonstrate switching between these quantum topological regimes and show that the process is robust to perturbations. Finally, we unveil possible preparation of a tripartite Greenberger-Horne-Zeilinger (GHZ) entanglement, distributing quantum correlations across polarization, OAM, and wavelength with only two photons. These results offer new routes to encode and manipulate multipartite topological entanglement in quantum optics.

Our core idea is illustrated conceptually in Fig.\,1(a). We start with a dual‑wavelength photon pair, entangled in both spatial (OAM) and spectral (wavelength) degrees of freedom, which then pass through a tunable, wavelength and voltage-dependent spin-orbit device. By adjusting a control parameter $p$ (here, an applied voltage) and applying projection schemes $\Pi_i^{\rm x}$ on each of the three degrees of freedom of photons $i=A,B$ with $\mathrm{x} \in \{\mathrm{OAM,\,polarization,\,wavelength}\}$, one can deterministically prepare (i) trivial two photon states with no topology (ii) dual-wavelength bipartite entangled skyrmion existing non-locally, see left panel of Fig.\,1(a) (iii) heralded single photon skyrmion, see right panel of Fig.\,1(a), and (iv) in principle, tripartite GHZ‑like entangled states spanning polarization, OAM, and wavelength.

\begin{figure*}[!ht]
\centering
\includegraphics[width=1\linewidth]{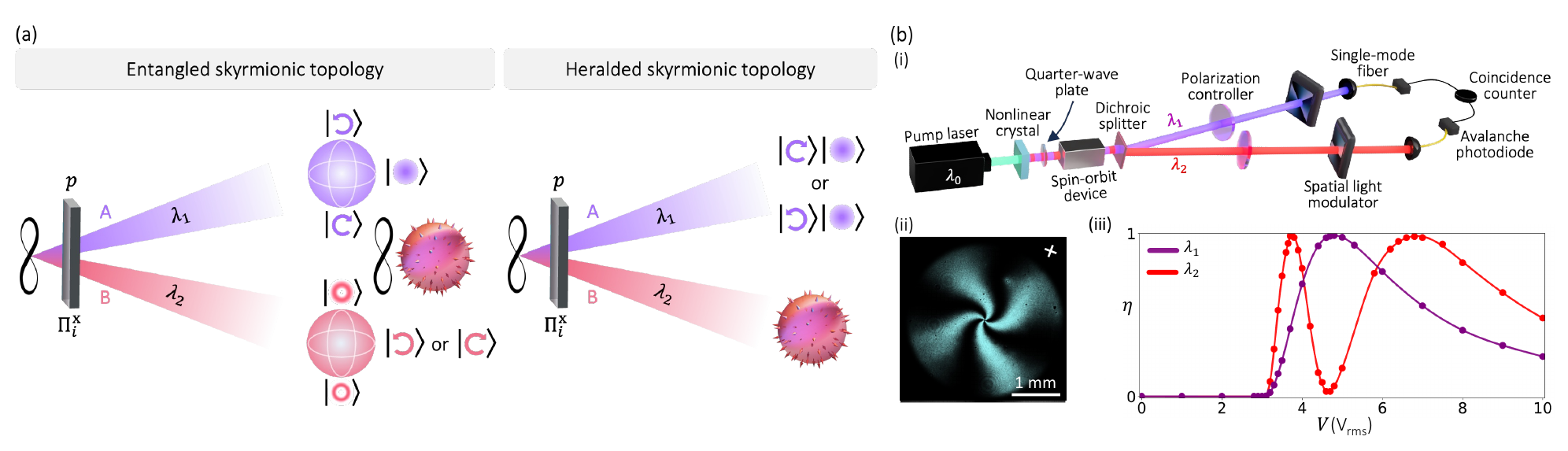}
\caption{(a) Illustration of the quantum topological switch. A dual-wavelength entangled photon pair described by Eq.\,\eqref{eq:initial_entangled_state} passes through a tunable, wavelength-dependent spin-orbit coupling device. By adjusting the control parameter $p$ (in this work, an applied voltage) and applying projection schemes $\Pi_i^{\rm x}$ on each of the three degrees of freedom of photons $i=\A,\B$ with ${\rm x} = {\rm (OAM, polarization, wavelength)}$, various quantum states can be prepared, including entangled skyrmionic topology and heralded single photons with a skyrmionic topology, with full details in the main text. (b)(i) Sketch of the experimental setup (see Supplemental Material Sec.\,IV for details \cite{Supplement}). (ii) Crossed linear polarizers image of the optical spin-orbit liquid crystal device at wavelength $\lambda_1$ and $V=4.7~{\rm V}$. (iii) Spin-to-orbital conversion efficiency vs applied voltage (all voltages are given in rms value). 
\label{fig:concept}
}
\end{figure*}

We now outline our scheme rigorously. Our implementation, depicted in Fig.\,1(b)(i), relies on the generation of collinear entangled pair of photons labeled A and B via type-0 non-degenerate spontaneous parametric down-conversion, pumped by a  continuous-wave laser at wavelength $\lambda_0 = 532$\,nm to produce entangled photons at wavelengths $\lambda_1 = 1550$\,nm and $\lambda_2 = 810$\,nm. Both photons emerge from the crystal with vertical polarization and are converted to right (R) or left (L) circular polarization via a broadband quarter-wave plate. Prior to any spatial or spectral filtering, assuming R polarization without loss of generality, and expressing the transverse structure in the OAM basis, the two-photon state---constrained by angular momentum conservation---reads
\begin{equation}
\ket{\Psi} \!= \!\sum_{l \in \mathbb{Z}} \! \sum_{k=1,2} \!\!\!\ c_{l,k} \ket{l}_\A \! \ket{\rm R}_\A \!\ket{\lambda_k}_\A \!\ket{-l}_\B \! \ket{\rm R}_\B \! \ket{\bar\lambda_k}_\B,
\label{eq:initial_entangled_state}
\end{equation}
where $c_{l,k}$ is the joint amplitude for detecting photon A with OAM $l\hbar$ at $\lambda_k$, and photon B with OAM $-l\hbar$ at $\bar\lambda_k$ satisfying $1/\lambda_k + 1/\bar\lambda_k = 1/\lambda_0$.

Spin-orbit optical engineering is applied using a dual-wavelength nematic liquid crystal q-plate \cite{yan2015q}, realised by the magnetoelectric generation of a unit charge {\it true} liquid crystal topological defect \cite{brasselet_prl_2018} shown in Fig.\,1(b)(ii). The device consists of 10 $\mu$m thickness nematic layer sandwiched between transparent electrodes, combined with a cylindrical neodymium ring magnet that induces an orientational defect with $q=1$. This structure provides an azimuthally varying optical axis and an electrically tunable birefringent retardation, enabling voltage-controlled spin–orbit conversion at both signal wavelengths. Fabrication and stability details are given in Supplemental Material Sec.~I \cite{Supplement}. This yields, at each wavelength, the transformation 
\begin{equation}
\ket{l}_i \!\ket{\rm R}_i  \to  \sqrt{1 - \eta} \ket{l}_i \! \ket{\rm R}_i + \sqrt{\eta} \ket{l - 2}_i \!\ket{\rm L}_i,
\label{eq:spinorbit_engineering}
\end{equation}
where $i = \A,\B$ and $\eta$ denotes the spin-to-orbital conversion efficiency of the device, which depends on both wavelength and applied voltage $V$, see Fig.\,1(b)(iii). Since any value in the range $0 < \eta < 1$ will suffice for our approach, we henceforth remove prefactors involving $\eta$ and instead present the states in unnormalised form.

Our approach relies on removing the spectral ambiguity from the initial entangled state $\ket{\Psi}$, without disturbing its spatial correlations. This is achieved using a dichroic beam splitter that separates photons A and B into two arms, associated with wavelengths $\lambda_1$ and $\lambda_2$, as shown in Fig.\,1(b)(i). Based on this principle, we demonstrate the projection-controlled generation of distinct quantum states. For illustration, we consider the generic case where both photons A and B undergo partial spin-to-orbital conversion, shown schematically in  Fig.\,1(b)(iii).  Following this photon A is coupled into a single-mode fiber, projecting it onto the $l = 0$ Gaussian mode. Although this operation acts only on photon A, it conditions the OAM state of photon B through their initial entanglement. The resulting two-photon state is (see Supplemental Material Sec.~II and Sec.~III for the derivation in the general case \cite{Supplement})
\begin{eqnarray}
\nonumber
\ket{\Psi'} & \propto & ( \ket{0}_\B \ket{\rm R}_\B + \ket{-2}_\B \ket{\rm L}_\B )  \ket{\rm R}_\A\\
&+&  ( \ket{-2}_\B \ket{\rm R}_\B +  \ket{-4}_\B \ket{\rm L}_\B ) \ket{\rm L}_\A ,
\label{eq:partial_conversion}
\end{eqnarray}
which enables the generation of distinct skyrmionic quantum states from polarization measurements.

Circular polarization measurement of photon B yields two options depending on R or L choice. Namely,
\begin{eqnarray}
\label{eq:nonlocal_R}
\hspace{-4mm} \ket{\Psi_{{\rm nonlocal},\rm R}}  & \propto & \ket{0}_\B \ket{\rm R}_\A   +\ket{-2}_\B  \ket{\rm L}_\A ,\\
\label{eq:nonlocal_L}
\hspace{-4mm} \ket{\Psi_{{\rm nonlocal},\rm L}} & \propto &  \ket{-2}_\B \ket{\rm R}_\A  + \ket{-4}_\B \ket{\rm L}_\A ,
\end{eqnarray}
where the OAM of photon B is entangled with the polarization of photon A. These correlations define an {\it entangled skyrmionic topology}, with the topological structure distributed nonlocally between spatial and polarization degrees of freedom across the two photons with distinct wavelengths defining a multi-colour skyrmion.

Alternatively, a polarization measurement on photon A in the circular polarization basis yields
\begin{eqnarray}
\label{eq:local_R}
\ket{\Psi_{{\rm local},\rm R}} & \propto &  \ket{0}_\B \ket{\rm R}_\B + \ket{-2}_\B \ket{\rm L}_\B,\\
\label{eq:local_L}
\ket{\Psi_{{\rm local},\rm L}}  & \propto & \ket{-2}_\B \ket{\rm R}_\B +  \ket{-4}_\B \ket{\rm L}_\B,
\end{eqnarray}
defining an {\it heralded skyrmionic topology} encoded locally in the polarization and spatial structure of photon B at a defined colour.

The use of single-mode fibers in both arms provides a way to manipulate the entanglement structure. Projecting photon B (rather than A) onto $l=0$ enables (i) entanglement of photon A's OAM with photon B's polarization via polarization measurement of A, and (ii) heralding of a local skyrmionic topology onto photon A via polarization measurement of B. Moreover, tuning the spin-orbit device to zero or full spin-to-orbital conversion for photon B (A), while photon A (B) is projected onto $l=0$, allows generation of entangled skyrmionic topologies without polarization post-selection. In contrast, applying full conversion to photon A (B) while simultaneously projecting it onto $l=0$ suppresses entanglement, yielding a state with trivial topology.

We first confirm the entangled nature of the generated photon pairs in the $l = \pm 2$ OAM subspace returning a concurrence of $C = 0.996$ and a CHSH Bell parameter of $S = 2.43$, thereby validating the source and enabling the experimental demonstration of the topological  (see Supplementary Sec.~V \cite{Supplement}). With the spin-orbit device in place and the voltage set to 3.85~V for obtaining a non-local state, we find a concurrence of $C = 0.83$, exceeding the entanglement threshold of $0.7$ and corresponding to a minimum expected CHSH Bell parameter \cite{verstraete2002entanglement} of $S = 2.34$. The drop in performance can be attributed to the lossy nature of the device, reducing the coincidences to $\approx 10\%$ of the original, but this could be improved in future engineered versions. Nevertheless, the measurements confirm that entanglement is preserved, with full details provided in Supplementary Material, Sec.~VII \cite{Supplement}. 

Next we fully unravel the quantum state with the aid of a quantum state tomography \cite{toninelli2019concepts} to experimentally reveal the nonlocal and local skyrmionic topologies. Polarization projections were carried out using suitable combinations of quarter-wave plates, half-wave plates, and linear polarizers to select the standard basis states: right (R), left (L), horizontal (H), vertical (V), diagonal (D), and anti-diagonal (A). OAM projections used forked holograms displayed on spatial light modulators, as in the initial entanglement verification, but here targeting either $\ket{-2}$, $\ket{-4}$, or one of the four equal-weight superpositions $(\ket{-2} + e^{i\phi} \ket{-4})/\sqrt{2}$ with $\phi = 0, \pi/2, \pi, 3\pi/2$. The results of these measurements are shown in Fig.\,2(a) for both the dual-wavelength entangled topology and the single-wavelength heralded topologies. We emphasize that these results correspond to raw data, with no subtraction of background or accidental counts.

\begin{figure}[b!]
\centering
\includegraphics[width= \linewidth]{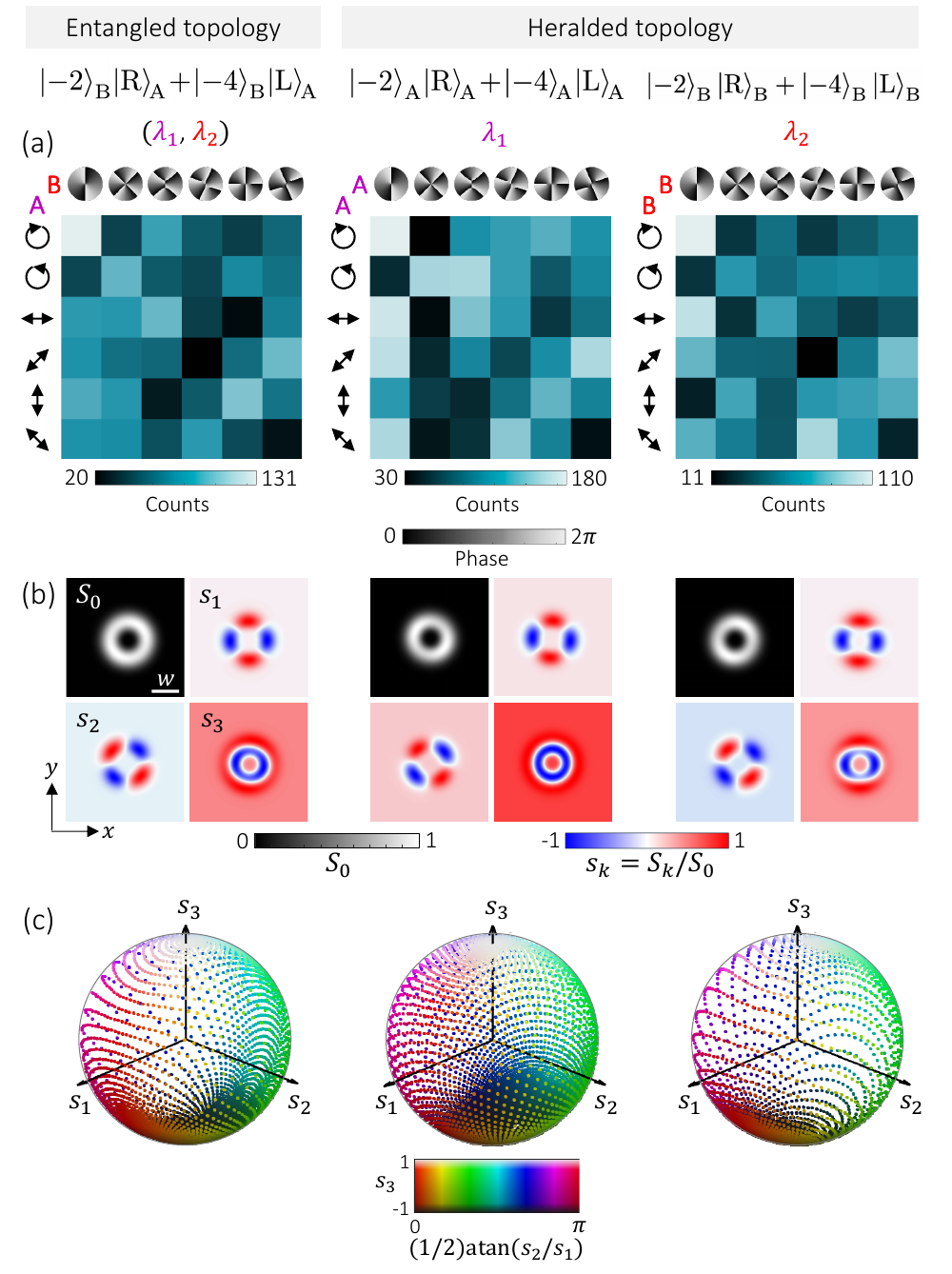}
\caption{Experimental observation of dual-wavelength nonlocal and single-wavelength local quantum skyrmionic topologies associated to the three main typical states under investigation. (a) Spin-orbit quantum state tomography based on combined polarization and OAM projections. Thumbnails to the left of each row indicate the polarization basis; those above each column show the spatial phase profiles of the projected state. (b) Transverse spatial distribution of the reduced Stokes vector components. (c) Corresponding Stokes vector textures mapped onto the Poincaré sphere, revealing skyrmionic polarization patterns with Skyrme number $N \approx -2$ in all cases. Experimental parameters: $V = 3.9$\,V for the entangled topology; $V = 6.3$\,V and $5.4$\,V for the heralded topologies at $\lambda_1$ and $\lambda_2$, respectively.}
\label{fig:QST_topologies}
\end{figure}

This tomographic data allows determining the $4 \times 4$ spin-orbit density matrix $\rho$ of either two-photon or single-photon states (see Supplementary Material, Sec.~VI \cite{Supplement},  for reconstructed density matrices), from which the spatially resolved quantum Stokes vector ${\bf S} = (S_1, S_2, S_3)$, with norm $S_0$, can be extracted \cite{ornelas2024non}.  In practice, we first project the full spin-orbit state $\rho$ onto a spatial mode basis $f_n(x,y)$ of Laguerre–Gauss functions, $f_n(r)\;\propto\;(r/w)^{|l_n|}\,\exp\bigl(-r^2/w^2\bigr) \exp\bigl(il_n\phi\bigr)$, to obtain the reduced $2\times2$ polarization density matrix $\bar\rho(x,y)\;=\;\operatorname{Tr}_{\rm spatial}\bigl[\rho\,(P_{x,y}\otimes\mathbb I_{\rm pol})\bigr]$, where $P_{x,y}=\ket{x,y}\bra{x,y} $ projects onto the transverse position  $(x,y)$ and  $ \mathbb I_{\rm pol}$ is the identity in the polarization subspace. We then compute Stokes parameter via $S_0(x,y)=\operatorname{Tr}\bigl[\bar\rho(x,y)\bigr],$ $S_k(x,y) = \operatorname{Tr}\left[\bar{\rho}(x,y) \sigma_k \right]$, $ k=1,2,3$, which expresses the local Stokes vector components as projections of $\rho$ onto the spatial position basis and the polarization Pauli operators $\sigma_k$. Expanding the density matrix in the OAM–polarization product basis, $\rho= \sum_{m,n,a,b} \tau_{mnab}\ket{f_m, e_a }\bra{f_n, e_b},$ where $\ket{f_m}$ denotes spatial modes, $\ket{e_a}$ $\in$ $\{\ket{R},\ket{L} \}$ the polarization basis, and $\tau_{mnab}$ the corresponding expansion coefficients. Noting $\innerp{x,y}{f_m}=f_m (x,y)$, we can express the spatially resolved Stokes components explicitly as $ S_k(x,y) = \sum_{m,n,a,b} \tau_{mnab} \, f_m(x,y) \, f_n^*(x,y) \, \operatorname{Tr}\left[\sigma_k \ket{e_a}\bra{e_b} \right] $. The resulting maps of ${\bf S}(x,y)$ are shown in Fig.\,2(b), with additional data provided in Supplemental Material, Sec.\,VIII \cite{Supplement}.

Each quantum Stokes vector field defines a mapping from the transverse real space to the Poincaré sphere of polarization, characterizing the local polarization expectation. To quantify the topology of these polarization textures, we evaluate the Skyrme number of the normalized vector field ${\bf s} = {\bf S}/S_0$,
\begin{equation}
N = \frac{1}{4\pi} \iint \mathbf{s} \cdot \left( \frac{\partial \mathbf{s}}{\partial x}\times \frac{\partial \mathbf{s}}{\partial y} \right)\,dxdy\,,
\label{eq:N}
\end{equation}
which counts how many times ${\bf s}$ wraps the Poincaré sphere. We find $N \approx -2$ with less than 1\% deviation in all cases, experimentally confirming the presence of second-order quantum skyrmions for both dual-wavelength entangled photon pairs and heralded single photons. Notably, had the initial two-photon state in Eq.\,\eqref{eq:initial_entangled_state} been prepared with L polarization instead of R, we would have observed oppositely charged skyrmions with $N = 2$.

\begin{figure}[b!]
\centering
\includegraphics[width=1\linewidth]{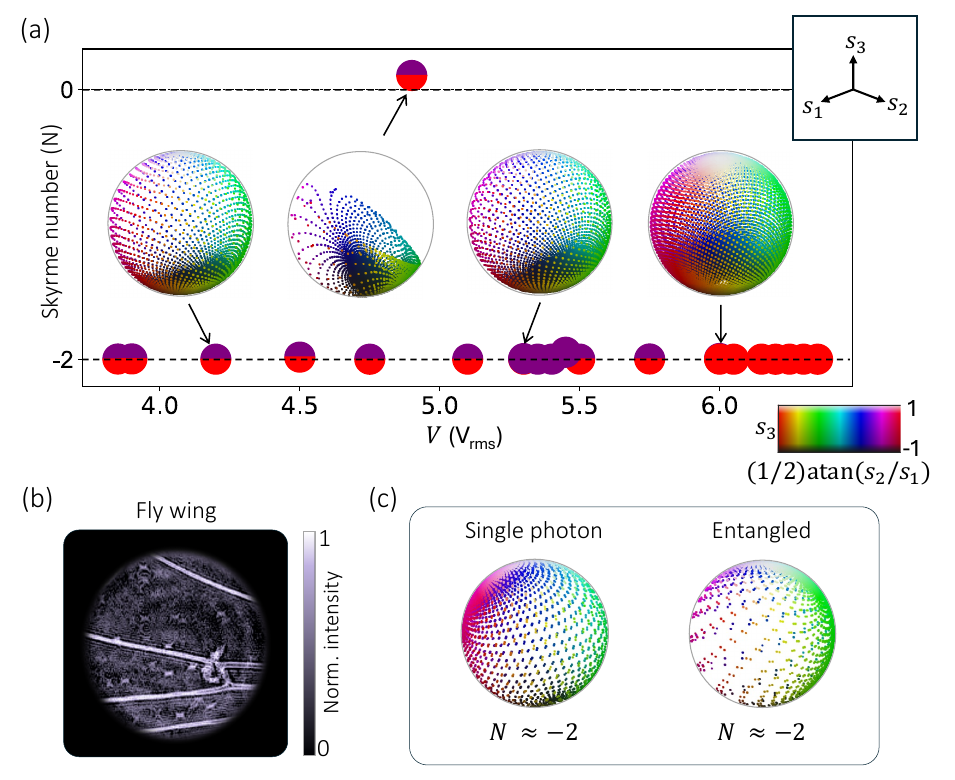}
\caption{(a) Skyrme number as a function of the voltage applied to the spin-orbit device, for different circular polarization projections. Depending on the projection, the measurement probes either dual-wavelength two-photon states (bicolor markers) or heralded single-photon states (monocolor markers: purple for $\lambda_1$, red for $\lambda_2$). Insets show typical Poincaré sphere coverage for representative cases at distinct voltages from that used in Fig.\,2: dual-wavelength entangled Skyrmion ($V=4.2$\,V), dual-wavelength trivial topology ($V=4.9$\,V), and single-wavelength heralded skyrmion ($V=6.0$\,V at $\lambda_1$ and $V=5.3$\,V at $\lambda_2$). (b) The generated quantum Skyrmion states are transmitted through a fly wing mounted to a microscope slide (pictured here as illuminated by a plane wave at wavelength, $\lambda=633$ nm) with (c) the resulting Poincaré sphere coverage for both single and entangled photon topological states.
}
\label{fig:fidelity}
\end{figure}

The ability to generate entangled and heralded topologies is not limited to the specific spin-orbit configurations shown in Fig.\,2, as further supported by our predictions. This is demonstrated in Fig.\,3, which shows the measured Skyrme number $N$ as a function of applied voltage for spin-orbit projections yielding either dual-wavelength two-photon states or heralded single-photon states. Notably, near $V = 4.9$\,V---the voltage maximizing spin-to-orbital conversion for photon A [Fig.\,1(b)(iii)] and corresponding to a trivial topology---we indeed measure $N \approx 0$, and a typical sparsely covered Poincar{\'e} sphere is shown as an inset in Fig.\,3(a). 

Notably, the generated topological states are robust against experimental imperfections and appear at wavelengths well-suited to bio-imaging. Indeed, the infrared photon enables minimally invasive probing of biological specimens, while the near-infrared photon can be efficiently detected with standard single-photon cameras. A first step in this direction is demonstrated through a proof-of-principle experiment, in which our entangled and single photon skyrmion states are transmitted through a fly wing shown in Fig.~\ref{fig:fidelity}(b). In both cases, the infrared photon was transmitted through the sample. We find that the topology remains intact at $N \approx -2$ for both single photon and entangled topological states, as seen in Fig.~\ref{fig:fidelity}(c) from the experimentally reconstructed topologies. The demonstrated preservation of the topology highlights its \textit{potential} for probing biological samples. 

Finally, we present a theoretical outlook suggesting that our experimental approach could, in principle, enable GHZ-like entanglement across three distinct degrees of freedom, which remains to be tested experimentally. Starting from the two-photon entangled state of Eq.\,\eqref{eq:initial_entangled_state}, entangled in both OAM and wavelength. Assuming full spin-to-orbital conversion at $\lambda_1$ while leaving the $\lambda_2$ state unaltered---as occurs under trivial topology when spectral ambiguity is lifted by the dichroic beamsplitter---we obtain our general six particle like GHZ state (see Supplemental Material, Sec.~IX \cite{Supplement})
\bea
\nonumber
\ket{\Psi} &=& \sum_{l\in \mathbb Z} \gamma_l \big (  \ket{\lambda_1}_A \ket{L}_A \ket{l - 2}_A \ket{\lambda_2}_B \ket{R}_B \ket{-l}_B + \\&&
\ket{\lambda_2}_A \ket{R}_A \ket{l}_A \ket{\lambda_1}_B \ket{L}_B \ket{-l - 2}_B\big )\,.
\label{eq:GHZ_preparation}
\eea
in which each photon encodes three logical qubits (OAM, polarization, wavelength). By projecting photon A first in spatial balanced superposition $\ket{+l}=\frac{\ket{l}+\ket{l-2}}{\sqrt{2}}$, then in spectral balanced superposition $\ket{+\lambda}=\frac{\ket{\lambda_1}+\ket{\lambda_2}}{\sqrt{2}}$ and projecting photon B into polarization superposition $\ket{+\sigma}=\frac{\ket{R}+\ket{L}}{\sqrt{2}}$, all information about the individual eigenstates is erased, coherently selecting only two terms and heralding the three-partite state
\begin{equation}
\ket{\Psi}_{\rm proj}\propto\ket{R}_A\ket{\lambda_1}_B\ket{-l-2}_B+\ket{L}_A\ket{\lambda_2}_B\ket{-l}_B.
\end{equation}

This state exhibits genuine GHZ entanglement across photon A’s polarization and photon B’s OAM and wavelength, and—under the mappings $(\ket{R}_A, \ket{L}_A) \rightarrow (\ket{0}, \ket{1})$, $(\ket{\lambda_1}_B, \ket{\lambda_2}_B) \rightarrow (\ket{0}, \ket{1})$ and $(\ket{-l-2}_B, \ket{-l}_B) \rightarrow (\ket{0}, \ket{1})$ is unitarily equivalent to the GHZ state $\frac{1}{\sqrt{2}}(\ket{0,0,0} + \ket{1,1,1})$. This distinguishes our demonstration from previous works on classical and quantum optical skyrmions, where wavelength appears as an independent, separable degree of freedom, i.e., $\ket{\Psi_{\A\B}} =\ket{\text{skyrmion}_{\A\B}} \otimes \ket{\lambda}$. In contrast, here the wavelength is nonseparable and contributes intrinsically to the skyrmionic structure of the entangled wavefunction: $\ket{\Psi_{\A\B}} \neq \ket{\text{skyrmion}_{\A\B}} \otimes \ket{\lambda}$.

In summary, we have demonstrated projective switching between skyrmion topologies in both two-photon entangled states and heralded single-photon states---realizations so far restricted to single-wavelength configurations and not extended to heralded photons. Our dual-wavelength approach further enables voltage-controlled generation of two-photon tripartite GHZ states entangled across polarization, orbital angular momentum, and wavelength. Moreover, the scheme is robust against experimental imperfections and operates at wavelengths relevant to fiber optics and bio-imaging, hence expanding the optical skyrmion toolbox and opening new avenues to explore with quantum topologies. From a materials standpoint, while patterned liquid crystal elements have been used in quantum optics since 2009 \cite{nagali2009quantum}, the present work uniquely leverages true liquid crystal defects, whose topological richness---so far exploited in classical structured light~\cite{brasselet_bookchapter_2021}---now enters the quantum regime.

\vspace{3mm}
The authors thank Pedro Ornelas and Neelan Gounden for useful discussions and resources. A. Forbes and I. Nape acknowledge funding from SA QuTI and E. Brasselet from IdEx University of Bordeaux/Grand Research Program GPR LIGHT. All authors acknowledge support from the CNRS-Wits collaboration fund.

\section{Supplementary Material}
\section{I Spin-orbit device}

\begin{figure*}[!ht]
    \centering
    \includegraphics[width=1\linewidth]{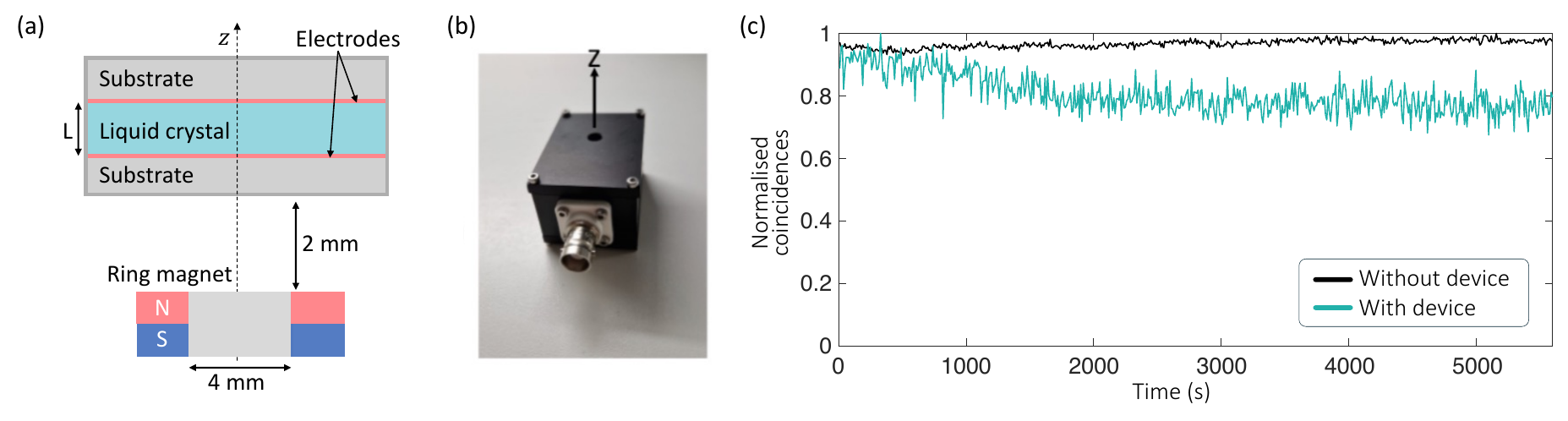}
    \caption{\textbf{Spin-orbit device.} (a) Schematic diagram of the device in which nematic liquid crystals are sandwiched between electrodes, and a ring magnet placed at distance $d$ = 2 mm. (b) Real image of packaged device highlighting the optical $z$ axis. (c) Coincidence counts with and without the spin orbit device monitored over $\sim1.5$ hours and normalised to the peak for each data set to allow for direct comparison. The measurements were taken over a 10s integration time to account for fluctuations and the stability with the device was monitored at a voltage of $V = 3.9$ V (by way of example) and without any alignment adjustments.}
    \label{fig:dev}
\end{figure*}


The spin-orbit device itself is made of a nematic liquid crystal (LC) cell and a cylindrical neodymium ring magnet (grade N50 with field strength $~1.40-1.46$ T). The LC cell (EHC Co. Ltd., Japan) is $10$ $\mathrm{\mu m}$ thick and filled with 1859 A nematic mixture (MUT, Poland) featuring $\Delta n \simeq 0.229$ birefringence at $\lambda = 589$~nm and negative dielectric anisotropy at 100~kHz frequency. Both inner substrates are coated with indium-tin-oxide electrodes for voltage control and treated to impose homeotropic anchoring, ensuring a uniform initial molecular alignment along the optical ($z$) axis. The ring magnet, with a height of 6~mm, outer diameter of 12~mm, and inner diameter of 4~mm, is aligned with its axis along $z$ and positioned approximately 2~mm above the substrate, as shown in the inset of Fig.~\ref{fig:dev} (a). The combined action of the electric and magnetic fields induces the spontaneous formation of a stable topological defect with $q=+1$ charge \cite{brasselet_prl_2018}. This defect features an electrically tunable uniform birefringent phase retardation $\Delta$ and an azimuthally varying optical axis oriented at an angle $\psi = q\varphi$ relative to the $x$-axis, where $\varphi$ is the polar angle.

The stability of the device was monitored over a time period of $\sim1.5$ hours at an applied voltage of $V = 3.9$ V, corresponding to the generation of an entangled skyrmion state. During this time, coincidence counts for a projection onto the generated state were recorded, as shown in Fig. \ref{fig:dev}(c). For comparison, we also measured coincidence counts without the device in the setup and observed a noticeable drop when the device was inserted, highlighting its influence on long-term stability. The device’s stability is influenced by internal elastic forces and by external factors such as temperature fluctuations and mechanical vibrations. To mitigate experimental drift, the applied voltage was temporarily switched off and then reapplied, followed by a slight realignment of the device.

\section{II SPDC Conservation}

In the SPDC process, a non-linear crystal with second-order susceptibility is used to down-convert a high-energy, pump photon into two lower-energy photons, obeying energy and momentum conservation. By virtue of energy conservation, we have:

\begin{equation}
    \hbar \omega_p = \hbar \omega_s + \hbar \omega_ii,
    \label{energycondition}
\end{equation}

where $\omega_{p,s,i}$ are the angular frequencies of the pump, signal and idler photons, respectively. For the degenerate case, $\omega_s = \omega_i$, where both photons share the same wavelength, and $\omega_s \neq \omega_i$ for the non-degenerate or dual wavelength case, where the photons have different wavelengths. This enables the generation of entangled photons with distinct wavelengths, e.g., $\lambda_s, \lambda_i = 810, 1550$ nm, provided the energy conservation condition is met.

\section{III State derivation}

We begin with a continuous‑wave pump at $\lambda_p = 532\,$nm driving non‑degenerate SPDC in a $\chi^{(2)}$ crystal.  Prior to any spatial and spectral (equivalently wavelength) filtering, the two‑photon state in transverse position and wavelength reads
\begin{eqnarray}
\nonumber
\ket{\Psi}
= \iint d^2r_A\,d^2r_B\;\Phi(r_A,r_B;\omega_A,\omega_B)\\\,
\ket{r_A,\omega_A}_A\,\ket{r_B,\omega_B}_B\;\ket{H}_A\,\ket{H}_B,
\end{eqnarray}
with $\omega_A + \omega_B = \omega_p$. For notational clarity we introduce central frequencies $\omega_1 =2\pi \frac{c}{\lambda_1}\;(\lambda_1 = 1550\text{ nm}), \; \text{and} \; \omega_2 = 2\pi \frac{c}{\lambda_2}\;(\lambda_2 = 810\text{ nm}), $  and label each spectral frequency state ($\ket{\omega_{AB}}$) by its equivalent wavelength state representation  $(\ket{\lambda_{AB}})$.  Furthermore, it can be assumed that the spectral components of the state, peaking at $\lambda_1$ and $\lambda_2$ are nonoverlapping, so that we can label the wavelength components of the state using $\ket{\lambda_{12}}$ where $\langle \lambda_1|\lambda_2 \rangle = 0$. We expand the state in the join OAM-wavelength basis by $\hat I = \sum_{l=-\infty}^\infty \sum_{k=1}^2  \ket{l,\lambda_k}\!\bra{l,\lambda_k}.$ Enforcing OAM conservation ($l_B = -l_A$) gives 
\begin{equation}
\ket{\Psi}
= \sum_{l,k\in\{1,2\}} C_{l,k }\;
  \ket{l,\lambda_k}_A\,\ket{-l,\lambda_{3-k}}_B\;
  \ket{H}_A\,\ket{H}_B,
\end{equation}

where the joint amplitudes $C_{l,k}$ retain both spatial and spectral correlations. It should be noted that this does not indicate the existence of any hybrid state in which the wavelength of one photon is coupled to the OAM of the other. Rather, the correlations are such that the OAM of photon A is coupled only with the OAM of photon B, while the wavelength of photon A is coupled only with the wavelength of photon B. A broadband quarter‑wave plate on each arm effects $\ket{H}\to\ket{R}$, yielding $\ket{\Psi}
= \sum_{l,k} C_{l,k}\; \ket{l,\lambda_k}_A\,\ket{-l,\lambda_{3-k}}_B\; \ket{R}_A\,\ket{R}_B.$ Only at the dichroic mirror do we separate the two wavelength bands, projecting onto $\ket{\lambda_1}_A\ket{\lambda_2}_B$. After this projection the state becomes
\begin{equation}
\ket{\Psi} = \sum_{l}c_{l}\, \ket{l,\lambda_1}_A\,\ket{-l,\lambda_2}_B\;
\ket{R}_A\,\ket{R}_B,
\end{equation}
with $c_l \equiv C_{l,1}$ and the spectral degree of freedom now factorized but the OAM entanglement intact. Both photons are then sent through a voltage‑tunable spin–orbit plate whose action on
$\ket{l,\lambda_j}\ket{R}$ is 
\begin{eqnarray}
\nonumber
\ket{l,\lambda_j}\ket{R}
\;\rightarrow\;
\sqrt{1-\eta(\lambda_j,V)}\,\ket{l,\lambda_j}\ket{R}\\
\;+\;\sqrt{\eta(\lambda_j,V)}\,\ket{l-2,\lambda_j}\ket{L},
\end{eqnarray}
Applying the spin–orbit transformation to both photons in $\ket{\Psi}$ results in a four-term superposition coupling OAM and polarization across both arms. After this both arms are coupled into single-mode fibres (SMFs), which project onto the fundamental Gaussian mode:

\begin{equation}
\ket{\mathrm{SMF}} = \sqrt{\frac{2}{\pi w_0^2}} \int d^2r\, e^{-r^2/w_0^2} \ket{r}
\; \xrightarrow{\text{OAM basis}} \;
\ket{0}.
\end{equation}
This projection filters out all modes except $l=0$ on the chosen arm, and  thereby conditions the OAM of the other photon through their initial entanglement.To see this explicitly, consider the two contributions on arm A after the spin–orbit plate: $\ket{l}_A\ket{R}_A$ and $\ket{l-2}_A\ket{L}_A$. Coupling into the SMF projects with $\bra{0}_A$, so the first term evaluates to $\innerp{0}{l}_A\,\ket{-l}_B$, which is nonzero only when $l=0$, conditioning photon B into $\ket{0}_B$. The second term evaluates to $\innerp{0}{l-2}_A\,\ket{-l}_B$, which is nonzero only when $l=2$, conditioning photon B into $\ket{-2}_B$. Thus, Gaussian projection on one arm not only selects the trivial $\ket{0}$ contribution but also heralds shifted modes on the partner photon, with the surviving terms tagged by the polarization of the projected photon.

Operating the device such that conversion is partial on both photons, we apply the spin–orbit transformation rules outlined in Eqn 4, and Gaussian projection on either arm, the two-photon state becomes.
\begin{eqnarray}
\nonumber
\ket{\Psi'} \propto 
\ket{\lambda_2}_B \ket{\lambda_1}_A (
\ket{R}_A(\ket{0}_B \ket{R}_B + \ket{-2}_B \ket{L}_B) +\\ 
\nonumber
\ket{L}_A (\ket{-2}_B \ket{R}_B + \ket{-4}_B \ket{L}_B)
).
\end{eqnarray}
A spin measurement on photon A (heralding either $\ket{R}$ or $\ket{L}$) collapses photon B into a single-photon vector mode. Crucially, since both photons pass through the same plate and are coupled into SMFs, we may choose to project either photon into the Gaussian mode. For instance, if we instead project photon B into $l $= 0, photon A inherits the hybrid spin–OAM structure.

\section{IV Experimental details }

We outline experimental details of the setup shown in Fig. 1 (b) (ii) of the main text to obtain our results. A collimated green diode laser of wavelength $\lambda_p = 532 $ nm, and power $350$ mW was sent to a temperature-controlled, Type-0 Periodically Poled Potassium Titanyl Phosphate (PPKTP) non-linear crystal of length $5$ mm. The pump polarization was converted to vertical polarization and the temperature of the crystal was optimized to 60 $^{\circ}$C to obtain collinear entangled photons of wavelengths, $\lambda_2 = 810 $ nm and $\lambda_1 = 1550 $ nm. After the crystal a quarter-wave plate (QWP) was used with the fast axis at an angle of $45^{\circ}$, converting the polarization from vertical to right-circular. The crystal plane was imaged onto the spin-orbit device  with a beam diameter, $w_D = 0.7$ mm.  After passing both photons through the device, a dichroic mirror was used to separate the entangled photons into two distinct arms, transmitting photons of wavelength $\lambda_2 = 810$ nm and reflecting the $\lambda_1 = 1550$ nm beam (shown as transmission in Fig.1 (b)(i) for simplicity). The plane of the device was imaged onto PLUTO-NIR and PLUTO-TELCO (HoloEye) spatial light modulators (SLMs) on each arm. To perform the necessary projections and state tomography, we employed two waveplates and a spatial light modulator (SLM). In the non-local case, separate projections were carried out on each arm: an SLM was used to project onto the required OAM states on one arm, while a combination of a quarter-wave plate, half-wave plate, and linear polarizer was used on the other arm to project onto various polarization states. Due to the polarization sensitivity of SLMs, the light emerging from the SLM was horizontally polarized, allowing the SLM to function as a linear polarizer and serve as a filter for the polarization projections. The half-wave plate was rotated to access linear polarization states, and the quarter-wave plate enabled projection onto circular polarization states. In the local case, with the single-photon spin-orbit coupled state, both polarization and OAM degrees of freedom were measured locally on a single arm. After the SLMs, the photons from each arm were coupled to optical single-mode fibers (SMFs) connected to avalanche photo-diodes (APDs) for photon detection. The photon counting device (CC) measured the coincidences within a 3 ns detection window, detecting a maximum coincidence count rate of $\sim$200 counts/s.

\section{V Spatial entanglement verification without spin-orbit device}

\begin{figure}[ht!]
\centering
\includegraphics[width=1\columnwidth]{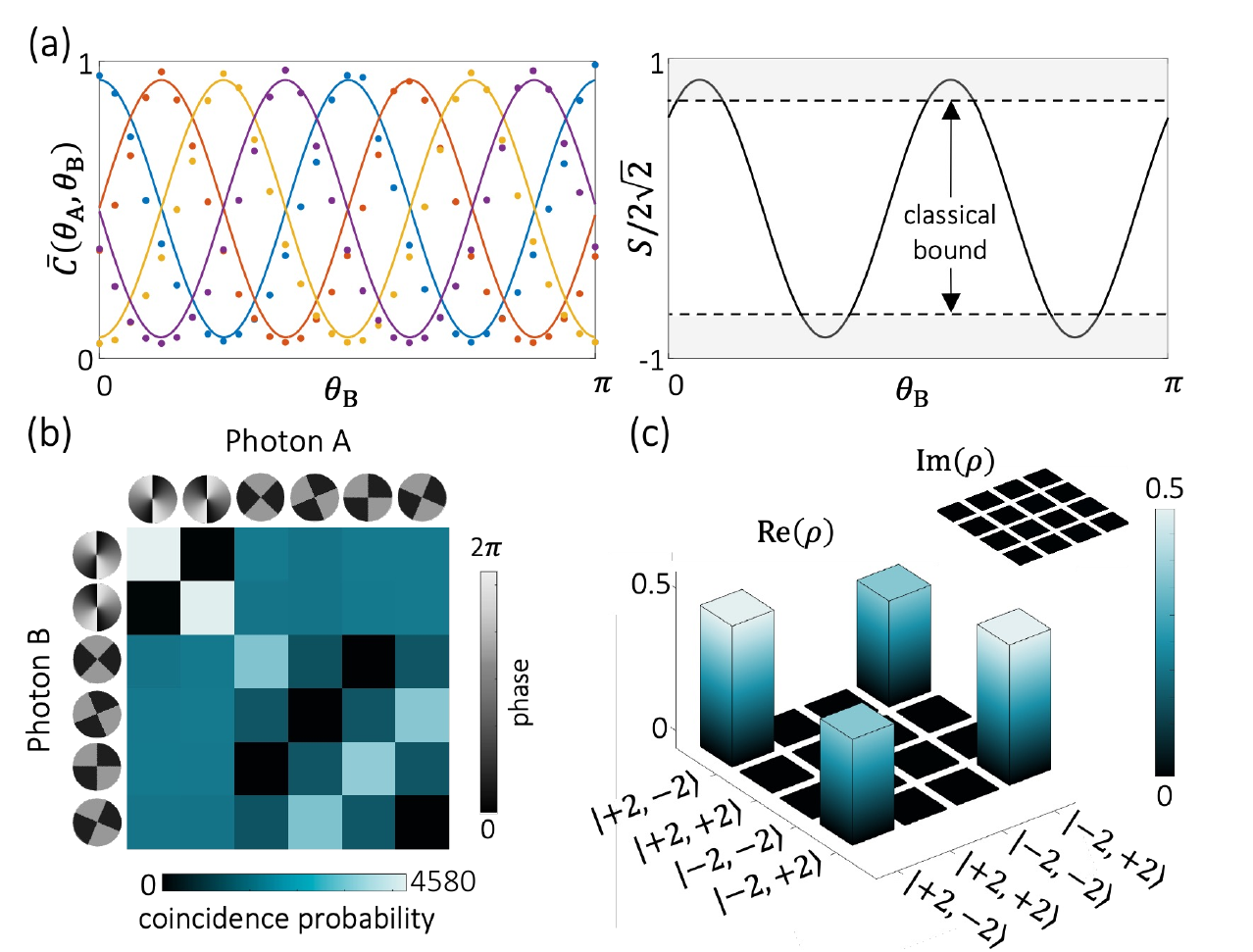}
\caption{Demonstration of OAM quantum entanglement of the prepared two-photon state in the absence of spin-orbit coupling. (a) Evaluation of the CHSH-Bell parameter $S$ from joint probabilities $C(\theta_A, \theta_B)$, with $\bar C = \max_{\theta_A,\theta_B}(C)$, for $\theta_A = 0$ (blue), $\tfrac{\pi}{8}$ (brown), $\tfrac{\pi}{4}$ (orange), and $\tfrac{3\pi}{8}$ (purple). (b) Quantum state tomography, where the projected states of each photon are identified by thumbnails depicting the corresponding spatial phase profiles. (c) Computed real and imaginary components of the density matrix from the data shown in panel (b). }
\label{fig:OAM_entanglement}
\end{figure}

Before probing the predicted nonlocal, local, and trivial topological regimes, we first verified OAM entanglement of the photon pairs generated by the nonlinear crystal. To this end, the spin-orbit device was removed and projective measurements were performed using suitable forked holograms on spatial light modulators in each arm, following the standard procedure \cite{leach2009violation} . This allowed evaluation of the Clauser-Horne-Shimony-Holt (CHSH) Bell parameter $S$ from coincidence counts $C(\theta_A, \theta_B)$ corresponding to the joint detecting probability of photon A  and B in the superpopisition states $\ket{\theta_{A(B)}}=\frac{1}{\sqrt{2}}\bigl(\ket{+2}+e^{il\theta_{A(B)}}\ket{-2}\bigr)$ , within the  OAM subspace $l=\pm 2$. Using $\theta_A \in \{0, \frac{\pi}{8}, \frac{\pi}{4}, \frac{3\pi}{8}\}$ and scanning $\theta_B$ over $[0,\pi]$, we measure $\max(S) = 2.43 $ as shown in Fig.\,2(a), which exceeds the classical limit $|S| = 2$ and approaches the upper quantum bound $|S| = 2\sqrt{2}$, thus confirming OAM entanglement.

The level of OAM entanglement is further assessed via quantum state tomography in this subspace, focusing on the target symmetric maximally entangled state $(|{-2}\rangle_A |{+2}\rangle_B + |{+2}\rangle_A |{-2}\rangle_B)/\sqrt{2}$. The protocol involves six projective OAM measurements per photon, defined by $\ket{-2}$, $\ket{2}$, and the four equal-weight superpositions $(\ket{-2} + e^{i\phi} \ket{+2})/\sqrt{2}$ with $\phi = 0, \pi/2, \pi, 3\pi/2$, resulting in 36 independent coincidence counts, as shown in Fig.\,2(b). We reconstruct the two‐photon density matrix from these measurements by expanding in the Gellmann operator basis,
\begin{equation}
\rho \;=\;\frac{1}{4}\,I_4 \;+\;\sum_{m,n=1}^3 b_{mn}\;\lambda^{(A)}_m\otimes\lambda^{(B)}_n,
\end{equation}
with $\{\lambda\}$ the standard Gellmann matrices and $b_{mn}$ the expansion coefficients.  These coefficients are obtained by minimizing the least‐squares cost function
\begin{equation}
\chi^2=\sum_{i,j}\bigl[p_{ij}-\operatorname{Tr}\{M_{ij}\,\rho\}\bigr]^2
\end{equation}
subject to $\rho\ge0$ and $\operatorname{Tr}\{\rho\}=1$ (enforcing $\frac{G^{\dagger}G}{Tr\{G^{\dagger}G\}}$ \cite{agnew2011tomography}, here G represents the density matrix elements in the Gellmann basis), with the measurements normalized in computational and mutually unbiased basis.  The reconstructed $4 \times 4$ density matrix $\rho$ provides full characterization of the generated quantum state, see Fig.\,2(c), and the fidelity $0 \leq F \leq 1$, quantifying its similarity to the target state associated with density matrix $\rho_{\rm t}$, yields $F = [{\rm Tr}(\sqrt{\sqrt{\rho_{\rm t}} \rho \sqrt{\rho_{\rm t}}})]^2 = 0.998$ and confirms a high degree of entanglement.

\section{VI Spin-orbit density matrices}

\begin{figure*}[!ht]
    \centering
    \includegraphics[width=1\linewidth]{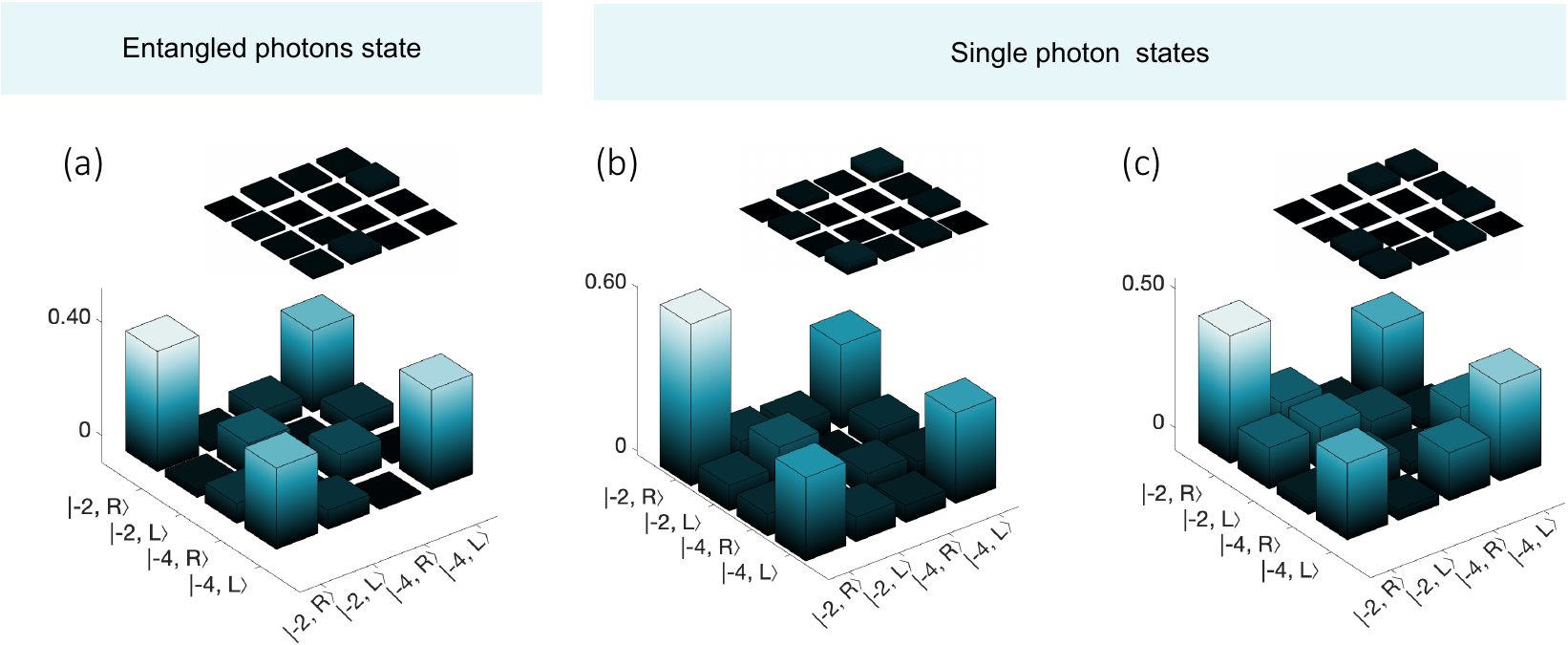}
    \caption{\textbf{Reconstructed spin-orbit density matrices}(a) The real and imaginary parts of reconstructed density matrices for all three cases as $4 \times 4$ bar plots}
    \label{fig:tomographh}
\end{figure*}

To reconstruct the spin–orbit density matrices utilized in the main text to extract spatially varying stokes parameters, we performed hybrid two‐photon quantum state tomography. In the entangled regime, photon A ($\lambda_1$ = 1550 nm) is projected on to six polarization eigenstates  $\{R,L,H, D, V, A\}$, while photon B ($\lambda_2$ = 810 nm) is measured in six spatial-OAM basis defined by  $\{\ket{-2}, \ket{-4}\}$ and their coherent superpositions $ \frac{1}{\sqrt{2}}(\ket{-2} \pm e^{i\phi} \ket{-4})$, with $\phi = 0, \frac{\pi}{2}, \pi, \frac{3\pi}{2}$. Altogether this yields $6\times6=36$ joint projectors $M_{ij}=P^{(A)}_i\otimes P^{(B)}_j$, where $P^{(A)}_i$ and $P^{(B)}_j$ act on photon A’s polarization and photon B’s spatial subspaces, respectively. For each projector we record the coincidence counts $C_{ij}$ (3 ns window) and normalize to probabilities for each projection basis $p_{ij} \;=\;\frac{C_{ij}}{\sum_{i,j}C_{ij}}$. In the heralded single photon regime (we imprint the topology on either on A  on B),one photon is first heralded in the $\ket{L}$ polarization state; an identical $6\times6$ tomography is then performed on its partner, again yielding normalized probabilities $p_{ij}$. From these measurements we reconstruct our density matrices shown Fig. 3 using the scheme outlined by eqn 2

\section{VII Post-device entanglement assessment}
Full two-qubit density matrices $\rho$ were reconstructed from measured coincidence counts using maximum-likelihood estimation (MLE), enforcing physicality (positivity and unit trace). From each reconstructed physical density matrix we computed the following standard figures of merit- concurrence, fidelity (defined Sec IV), purity as follows. The concurrence is computed following Wootters, start by forming the spin-flipped matrix $\tilde{\rho}=(\sigma_y\otimes\sigma_y)\,\rho^{*}\,(\sigma_y\otimes\sigma_y)$, evaluate the eigenvalues $\{\lambda_i\}$ of $\rho\tilde{\rho}$ ordered so that $\lambda_1\ge\lambda_2\ge\lambda_3\ge\lambda_4$, and set $C=\max\big(0,\sqrt{\lambda_1}-\sqrt{\lambda_2}-\sqrt{\lambda_3}-\sqrt{\lambda_4}\big)$.  Purity is defined as $P=\mathrm{Tr}(\rho^2)$ (with $P=1$ for pure states and $P<1$ for mixed states). No CHSH Bell test was performed with the LC device in place, instead concurrence is used as an entanglement witness and to bound the possible CHSH parameter. Following Verstraete and Wolf,  define $\beta$ by $S=2\beta$; for a given concurrence $C$ one has $\beta_{\min}(C)=\max\!\big(1,\sqrt{2}\,C\big)$ and $\beta_{\max}(C)=\sqrt{1+C^2}$, hence the CHSH parameter satisfies $S_{\min}(C)=2\cdot\max\!\big(1,\sqrt{2}\,C\big)$ and $S_{\max}(C)=2\sqrt{1+C^2}$.For example, the post-device concurrence reported in the main text, $C\approx0.83$, yields $S\in[2.34,\,2.60]$, i.e. values strictly above the local realistic bound $S=2$ in principle; we note that in practice experimental losses and detection limitations can reduce $S$ below the theoretical interval.

\begin{table}[ht!]
\centering
\label{tab:post_device_metrics}
\resizebox{\linewidth}{!}{
\begin{tabular}{c c c c c c}
\hline
Voltage (V) & Concurrence $C$ & $S_{\min}$ & $S_{\max}$ & Fidelity $F$ & Purity $P$ \\
\hline
3.85  & 0.8286 & 2.344 & 2.597 & 0.8920 & 0.8396 \\
3.9  & 0.7874 & 2.227 & 2.546 & 0.8735 & 0.8095 \\
4.2  & 0.6546 & 2.000 & 2.390 & 0.7934 & 0.8588 \\
4.5  & 0.1188 & 2.000 & 2.014 & 0.4999 & 0.5062 \\
4.75  & 0.3736 & 2.000 & 2.135 & 0.4138 & 0.5041 \\
4.9  & 0.0000 & 2.000 & 2.000 & 0.2516 & 0.5592 \\
5.1 & 0.3305 & 2.000 & 2.106 & 0.6293 & 0.6547 \\
5.3 & 0.6613 & 2.000 & 2.398 & 0.7509 & 0.7957 \\
5.5 & 0.7646 & 2.163 & 2.518 & 0.7479 & 0.8296 \\
5.75 & 0.7744 & 2.190 & 2.530 & 0.7542 & 0.8455 \\
6.0 & 0.8093 & 2.289 & 2.573 & 0.7271 & 0.9002 \\
\hline
\end{tabular}}
\caption{Tomography-derived metrics for post-device entangled hybrid states at different LC voltages. Concurrence $C$, fidelity $F$, and purity $P$ are obtained from maximum-likelihood reconstructed density matrices. The predicted CHSH interval $S_{\min}$--$S_{\max}$ follows Verstraete \& Wolf ; see text for formulae.}
\end{table}

\section{VIII Additional Poincaré sphere mappings}
To complement the Skyrme number analysis presented in the main text, we show here the evolution of the reconstructed polarization fields $\mathbf{S}(x,y)$ projected onto the Poincaré sphere at various sampled voltage points. Each sphere shows the spatially sampled Stokes vectors mapped to the Poincaré sphere, illustrating how the extracted polarization coverage changes with applied voltage- the control parameter of the spin-orbit device. As seen, broad and dense coverage of the Poincaré sphere corresponds to the topological regimes ($N \approx -2$), whereas limited coverage and polarization collapse signal the trivial regime ($N \approx 0$). 
\begin{figure*}[!ht]
    \centering
    \includegraphics[width=1\linewidth]{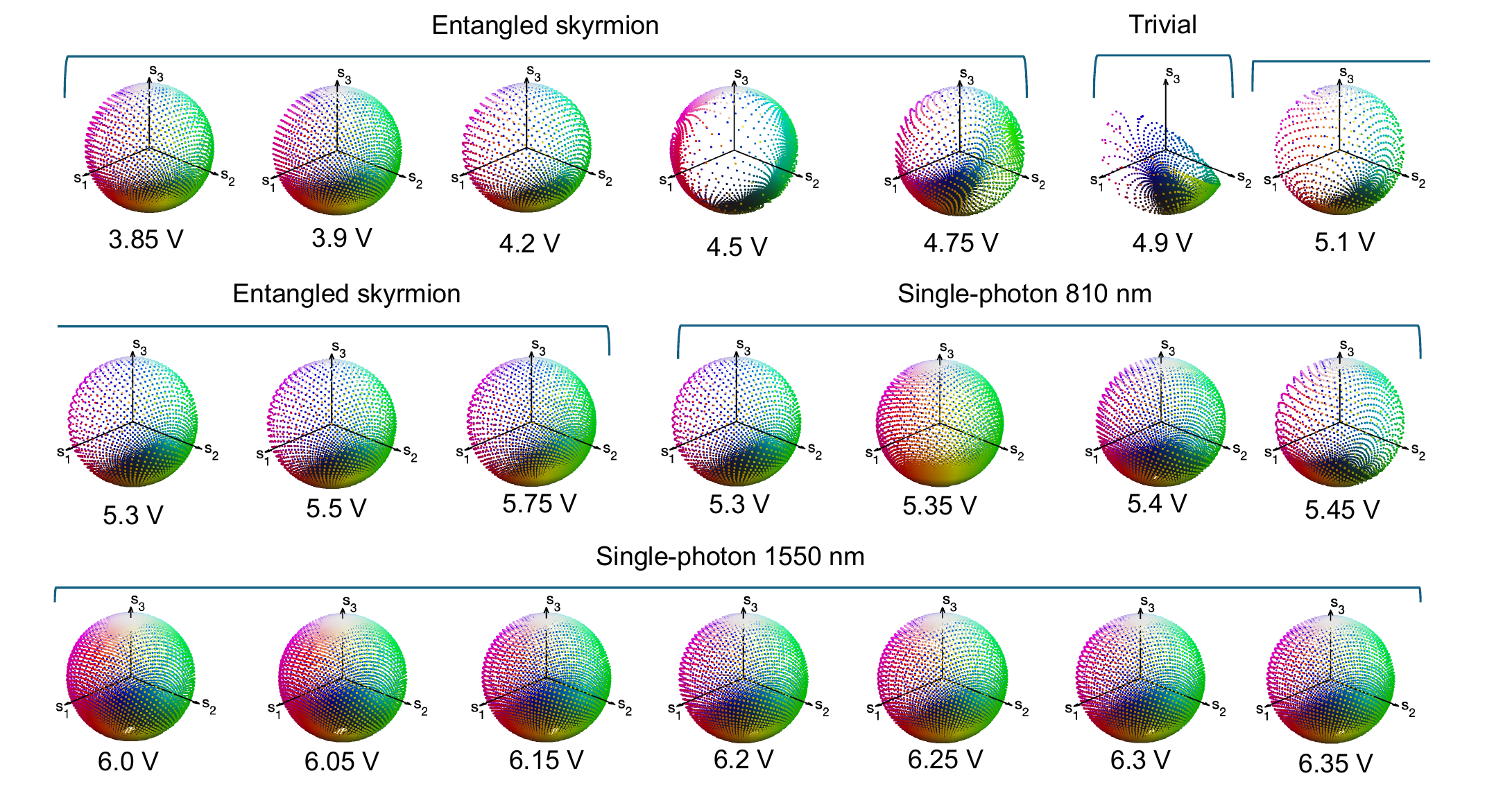}
    \caption{{Poincaré sphere representations of the polarization field $\mathbf{S}(x,y)$ at various applied voltages} Each sphere shows the spatially sampled Stokes vectors extracted from the reconstructed density matrices and projected onto the Poincaré sphere. The voltage sweep spans the different quantum topological regimes extracted in the experiment: entangled skyrmion (top rows), single-photon skyrmion at 810 nm and 1550 nm (middle and bottom rows), and the trivial topology region (centered around 4.9 V). Dense and near-complete coverage of the sphere indicates robust topological structure ($N \approx -2$), while sparse or localized distributions correspond to trivial or partially degraded topology.
    \label{fig:Stokes}}
\end{figure*}

\section{IX GHZ-like state derivation}
To illustrate our proposal for emulating GHZ-like entangled states across multiple degrees of freedom (DoFs), we begin with a two-photon state entangled in both orbital angular momentum (OAM) and wavelength:

\begin{equation}
\sum_{l,\lambda} c_{l,\lambda} \ket{l}_A \ket{\lambda_1}_A \ket{-l}_B \ket{\lambda_2}_B \ket{R}_A \ket{R}_B.
\end{equation}
Here, photons $A$ and $B$ are initially in right-circular polarization $R$ and share entanglement in OAM $\pm l$  and wavelength ($\lambda_1,\lambda_2$). We restrict ourselves to a two dimensional subspace in both OAM and wavelength, rewriting our state as

\begin{equation}
\begin{split}
    \ket{\Psi} = \frac{1}{\sqrt{2}} \Big(
\ket{l}_A \ket{\lambda_1}_A \ket{-l}_B \ket{\lambda_2}_B \ket{R}_A \ket{R}_B + \\ 
\ket{-l}_B \ket{\lambda_1}_B \ket{l}_A \ket{\lambda_2}_A \ket{R}_B \ket{R}_A \Big).
\end{split}
\end{equation}

Next, we transform this state by applying a spin-orbit transformation plate that operates in a wavelength-selective manner. Specifically, for photons at wavelength $\lambda_1$,  the plate induces the transformation $\ket{R, l} \rightarrow \ket{L, l - 2}$, while leaving photons at $\lambda_2$ unaffected. Under this transformation, the state becomes: 

\begin{equation}
\begin{split}
\ket{\Psi} = \frac{1}{\sqrt{2}} \Big(\ket{\lambda_1}_A \ket{L}_A \ket{l - 2}_A \ket{\lambda_2}_B \ket{R}_B \ket{-l}_B + \\
\ket{\lambda_2}_A \ket{R}_A \ket{l}_A \ket{\lambda_1}_B \ket{L}_B \ket{-l - 2}_B
\Big).
\end{split}
\end{equation}

Interestingly, at this operational point, the  wavelength ambiguity—normally filtered out by dichroic elements—serves as a resource, enabling  superposition of orthogonal spin components. The resulting state exhibits a structure analogous to a Greenberger–Horne–Zeilinger (GHZ) state, not across three particles, but across three degrees of freedom in each of two photons: wavelength, polarization, and OAM. We make this explicit by assigning logical qubits $\ket{0}_\lambda \rightarrow \ket{\lambda_1}$, $\ket{1}_\lambda \rightarrow  \ket{\lambda_2}$, $\ket{0}_\sigma \rightarrow \ket{R}$, $\ket{1}_\sigma\rightarrow  \ket{L}$,  $\ket{0}_l\rightarrow  \ket{l}$  and $\ket{1}_l\rightarrow  \ket{l-2}$ 
Expressed in this computational basis, the entangled state takes the form:
\begin{equation}
    \ket{\Psi}\propto\ket{0,1,1}_A\ket{1,0,0}_B +  \ket{1,0,0}_A\ket{0,1,1}_B
\end{equation}
where each triplet denotes the qubit state for ($\lambda,\sigma,l$) of photon A or B. While not a canonical GHZ state, this form exhibits the same essential feature: the entire state exists as a coherent superposition of two distinct configurations with  correlations between the subsystems.


\begin{thebibliography}{32}%
\makeatletter
\providecommand \@ifxundefined [1]{%
 \@ifx{#1\undefined}
}%
\providecommand \@ifnum [1]{%
 \ifnum #1\expandafter \@firstoftwo
 \else \expandafter \@secondoftwo
 \fi
}%
\providecommand \@ifx [1]{%
 \ifx #1\expandafter \@firstoftwo
 \else \expandafter \@secondoftwo
 \fi
}%
\providecommand \natexlab [1]{#1}%
\providecommand \enquote  [1]{``#1''}%
\providecommand \bibnamefont  [1]{#1}%
\providecommand \bibfnamefont [1]{#1}%
\providecommand \citenamefont [1]{#1}%
\providecommand \href@noop [0]{\@secondoftwo}%
\providecommand \href [0]{\begingroup \@sanitize@url \@href}%
\providecommand \@href[1]{\@@startlink{#1}\@@href}%
\providecommand \@@href[1]{\endgroup#1\@@endlink}%
\providecommand \@sanitize@url [0]{\catcode `\\12\catcode `\$12\catcode `\&12\catcode `\#12\catcode `\^12\catcode `\_12\catcode `\%12\relax}%
\providecommand \@@startlink[1]{}%
\providecommand \@@endlink[0]{}%
\providecommand \url  [0]{\begingroup\@sanitize@url \@url }%
\providecommand \@url [1]{\endgroup\@href {#1}{\urlprefix }}%
\providecommand \urlprefix  [0]{URL }%
\providecommand \Eprint [0]{\href }%
\providecommand \doibase [0]{http://dx.doi.org/}%
\providecommand \selectlanguage [0]{\@gobble}%
\providecommand \bibinfo  [0]{\@secondoftwo}%
\providecommand \bibfield  [0]{\@secondoftwo}%
\providecommand \translation [1]{[#1]}%
\providecommand \BibitemOpen [0]{}%
\providecommand \bibitemStop [0]{}%
\providecommand \bibitemNoStop [0]{.\EOS\space}%
\providecommand \EOS [0]{\spacefactor3000\relax}%
\providecommand \BibitemShut  [1]{\csname bibitem#1\endcsname}%
\let\auto@bib@innerbib\@empty
\bibitem [{\citenamefont {Leslie}\ \emph {et~al.}(2009)\citenamefont {Leslie}, \citenamefont {Hansen}, \citenamefont {Wright}, \citenamefont {Deutsch},\ and\ \citenamefont {Bigelow}}]{leslie2009creation}%
  \BibitemOpen
  \bibfield  {author} {\bibinfo {author} {\bibfnamefont {L.}~\bibnamefont {Leslie}}, \bibinfo {author} {\bibfnamefont {A.}~\bibnamefont {Hansen}}, \bibinfo {author} {\bibfnamefont {K.}~\bibnamefont {Wright}}, \bibinfo {author} {\bibfnamefont {B.}~\bibnamefont {Deutsch}}, \ and\ \bibinfo {author} {\bibfnamefont {N.}~\bibnamefont {Bigelow}},\ }\href@noop {} {\bibfield  {journal} {\bibinfo  {journal} {Physical Review Letters}\ }\textbf {\bibinfo {volume} {103}},\ \bibinfo {pages} {250401} (\bibinfo {year} {2009})}\BibitemShut {NoStop}%
\bibitem [{\citenamefont {Smalyukh}\ \emph {et~al.}(2010)\citenamefont {Smalyukh}, \citenamefont {Lansac}, \citenamefont {Clark},\ and\ \citenamefont {Trivedi}}]{smalyukh2010three}%
  \BibitemOpen
  \bibfield  {author} {\bibinfo {author} {\bibfnamefont {I.~I.}\ \bibnamefont {Smalyukh}}, \bibinfo {author} {\bibfnamefont {Y.}~\bibnamefont {Lansac}}, \bibinfo {author} {\bibfnamefont {N.~A.}\ \bibnamefont {Clark}}, \ and\ \bibinfo {author} {\bibfnamefont {R.~P.}\ \bibnamefont {Trivedi}},\ }\href@noop {} {\bibfield  {journal} {\bibinfo  {journal} {Nature Materials}\ }\textbf {\bibinfo {volume} {9}},\ \bibinfo {pages} {139} (\bibinfo {year} {2010})}\BibitemShut {NoStop}%
\bibitem [{\citenamefont {Chen}\ \emph {et~al.}(2013)\citenamefont {Chen}, \citenamefont {Ackerman}, \citenamefont {Alexander}, \citenamefont {Kamien},\ and\ \citenamefont {Smalyukh}}]{chen2013generating}%
  \BibitemOpen
  \bibfield  {author} {\bibinfo {author} {\bibfnamefont {B.~G.-g.}\ \bibnamefont {Chen}}, \bibinfo {author} {\bibfnamefont {P.~J.}\ \bibnamefont {Ackerman}}, \bibinfo {author} {\bibfnamefont {G.~P.}\ \bibnamefont {Alexander}}, \bibinfo {author} {\bibfnamefont {R.~D.}\ \bibnamefont {Kamien}}, \ and\ \bibinfo {author} {\bibfnamefont {I.~I.}\ \bibnamefont {Smalyukh}},\ }\href@noop {} {\bibfield  {journal} {\bibinfo  {journal} {Physical review letters}\ }\textbf {\bibinfo {volume} {110}},\ \bibinfo {pages} {237801} (\bibinfo {year} {2013})}\BibitemShut {NoStop}%
\bibitem [{\citenamefont {Ge}\ \emph {et~al.}(2021)\citenamefont {Ge}, \citenamefont {Xu}, \citenamefont {Liu}, \citenamefont {Xu}, \citenamefont {Lin}, \citenamefont {Yu}, \citenamefont {Bao}, \citenamefont {Jiang}, \citenamefont {Lu},\ and\ \citenamefont {Chen}}]{ge2021observation}%
  \BibitemOpen
  \bibfield  {author} {\bibinfo {author} {\bibfnamefont {H.}~\bibnamefont {Ge}}, \bibinfo {author} {\bibfnamefont {X.-Y.}\ \bibnamefont {Xu}}, \bibinfo {author} {\bibfnamefont {L.}~\bibnamefont {Liu}}, \bibinfo {author} {\bibfnamefont {R.}~\bibnamefont {Xu}}, \bibinfo {author} {\bibfnamefont {Z.-K.}\ \bibnamefont {Lin}}, \bibinfo {author} {\bibfnamefont {S.-Y.}\ \bibnamefont {Yu}}, \bibinfo {author} {\bibfnamefont {M.}~\bibnamefont {Bao}}, \bibinfo {author} {\bibfnamefont {J.-H.}\ \bibnamefont {Jiang}}, \bibinfo {author} {\bibfnamefont {M.-H.}\ \bibnamefont {Lu}}, \ and\ \bibinfo {author} {\bibfnamefont {Y.-F.}\ \bibnamefont {Chen}},\ }\href@noop {} {\bibfield  {journal} {\bibinfo  {journal} {Physical Review Letters}\ }\textbf {\bibinfo {volume} {127}},\ \bibinfo {pages} {144502} (\bibinfo {year} {2021})}\BibitemShut {NoStop}%
\bibitem [{\citenamefont {Muelas-Hurtado}\ \emph {et~al.}(2022)\citenamefont {Muelas-Hurtado}, \citenamefont {Volke-Sep{\'u}lveda}, \citenamefont {Ealo}, \citenamefont {Nori}, \citenamefont {Alonso}, \citenamefont {Bliokh},\ and\ \citenamefont {Brasselet}}]{muelas2022observation}%
  \BibitemOpen
  \bibfield  {author} {\bibinfo {author} {\bibfnamefont {R.~D.}\ \bibnamefont {Muelas-Hurtado}}, \bibinfo {author} {\bibfnamefont {K.}~\bibnamefont {Volke-Sep{\'u}lveda}}, \bibinfo {author} {\bibfnamefont {J.~L.}\ \bibnamefont {Ealo}}, \bibinfo {author} {\bibfnamefont {F.}~\bibnamefont {Nori}}, \bibinfo {author} {\bibfnamefont {M.~A.}\ \bibnamefont {Alonso}}, \bibinfo {author} {\bibfnamefont {K.~Y.}\ \bibnamefont {Bliokh}}, \ and\ \bibinfo {author} {\bibfnamefont {E.}~\bibnamefont {Brasselet}},\ }\href@noop {} {\bibfield  {journal} {\bibinfo  {journal} {Physical Review Letters}\ }\textbf {\bibinfo {volume} {129}},\ \bibinfo {pages} {204301} (\bibinfo {year} {2022})}\BibitemShut {NoStop}%
\bibitem [{\citenamefont {Wang}\ \emph {et~al.}(2025)\citenamefont {Wang}, \citenamefont {Che}, \citenamefont {Cheng}, \citenamefont {Tong}, \citenamefont {Shi}, \citenamefont {Shen}, \citenamefont {Bliokh},\ and\ \citenamefont {Zi}}]{wang2025topological}%
  \BibitemOpen
  \bibfield  {author} {\bibinfo {author} {\bibfnamefont {B.}~\bibnamefont {Wang}}, \bibinfo {author} {\bibfnamefont {Z.}~\bibnamefont {Che}}, \bibinfo {author} {\bibfnamefont {C.}~\bibnamefont {Cheng}}, \bibinfo {author} {\bibfnamefont {C.}~\bibnamefont {Tong}}, \bibinfo {author} {\bibfnamefont {L.}~\bibnamefont {Shi}}, \bibinfo {author} {\bibfnamefont {Y.}~\bibnamefont {Shen}}, \bibinfo {author} {\bibfnamefont {K.~Y.}\ \bibnamefont {Bliokh}}, \ and\ \bibinfo {author} {\bibfnamefont {J.}~\bibnamefont {Zi}},\ }\href {\doibase https://doi.org/10.1038/s41586-024-08384-y} {\bibfield  {journal} {\bibinfo  {journal} {Nature}\ } (\bibinfo {year} {2025}),\ https://doi.org/10.1038/s41586-024-08384-y}\BibitemShut {NoStop}%
\bibitem [{\citenamefont {Tsesses}\ \emph {et~al.}(2018)\citenamefont {Tsesses}, \citenamefont {Ostrovsky}, \citenamefont {Cohen}, \citenamefont {Gjonaj}, \citenamefont {Lindner},\ and\ \citenamefont {Bartal}}]{tsesses2018optical}%
  \BibitemOpen
  \bibfield  {author} {\bibinfo {author} {\bibfnamefont {S.}~\bibnamefont {Tsesses}}, \bibinfo {author} {\bibfnamefont {E.}~\bibnamefont {Ostrovsky}}, \bibinfo {author} {\bibfnamefont {K.}~\bibnamefont {Cohen}}, \bibinfo {author} {\bibfnamefont {B.}~\bibnamefont {Gjonaj}}, \bibinfo {author} {\bibfnamefont {N.}~\bibnamefont {Lindner}}, \ and\ \bibinfo {author} {\bibfnamefont {G.}~\bibnamefont {Bartal}},\ }\href@noop {} {\bibfield  {journal} {\bibinfo  {journal} {Science}\ }\textbf {\bibinfo {volume} {361}},\ \bibinfo {pages} {993} (\bibinfo {year} {2018})}\BibitemShut {NoStop}%
\bibitem [{\citenamefont {Guti{\'e}rrez-Cuevas}\ and\ \citenamefont {Pisanty}(2021)}]{gutierrez2021optical}%
  \BibitemOpen
  \bibfield  {author} {\bibinfo {author} {\bibfnamefont {R.}~\bibnamefont {Guti{\'e}rrez-Cuevas}}\ and\ \bibinfo {author} {\bibfnamefont {E.}~\bibnamefont {Pisanty}},\ }\href@noop {} {\bibfield  {journal} {\bibinfo  {journal} {Journal of Optics}\ }\textbf {\bibinfo {volume} {23}},\ \bibinfo {pages} {024004} (\bibinfo {year} {2021})}\BibitemShut {NoStop}%
\bibitem [{\citenamefont {Gao}\ \emph {et~al.}(2020)\citenamefont {Gao}, \citenamefont {Speirits}, \citenamefont {Castellucci}, \citenamefont {Franke-Arnold}, \citenamefont {Barnett},\ and\ \citenamefont {G{\"o}tte}}]{gao2020paraxial}%
  \BibitemOpen
  \bibfield  {author} {\bibinfo {author} {\bibfnamefont {S.}~\bibnamefont {Gao}}, \bibinfo {author} {\bibfnamefont {F.~C.}\ \bibnamefont {Speirits}}, \bibinfo {author} {\bibfnamefont {F.}~\bibnamefont {Castellucci}}, \bibinfo {author} {\bibfnamefont {S.}~\bibnamefont {Franke-Arnold}}, \bibinfo {author} {\bibfnamefont {S.~M.}\ \bibnamefont {Barnett}}, \ and\ \bibinfo {author} {\bibfnamefont {J.~B.}\ \bibnamefont {G{\"o}tte}},\ }\href@noop {} {\bibfield  {journal} {\bibinfo  {journal} {Physical Review A}\ }\textbf {\bibinfo {volume} {102}},\ \bibinfo {pages} {053513} (\bibinfo {year} {2020})}\BibitemShut {NoStop}%
\bibitem [{\citenamefont {Du}\ \emph {et~al.}(2019)\citenamefont {Du}, \citenamefont {Yang}, \citenamefont {Zayats},\ and\ \citenamefont {Yuan}}]{du2019deep}%
  \BibitemOpen
  \bibfield  {author} {\bibinfo {author} {\bibfnamefont {L.}~\bibnamefont {Du}}, \bibinfo {author} {\bibfnamefont {A.}~\bibnamefont {Yang}}, \bibinfo {author} {\bibfnamefont {A.~V.}\ \bibnamefont {Zayats}}, \ and\ \bibinfo {author} {\bibfnamefont {X.}~\bibnamefont {Yuan}},\ }\href@noop {} {\bibfield  {journal} {\bibinfo  {journal} {Nature Physics}\ }\textbf {\bibinfo {volume} {15}},\ \bibinfo {pages} {650} (\bibinfo {year} {2019})}\BibitemShut {NoStop}%
\bibitem [{\citenamefont {Sugic}\ \emph {et~al.}(2021)\citenamefont {Sugic}, \citenamefont {Droop}, \citenamefont {Otte}, \citenamefont {Ehrmanntraut}, \citenamefont {Nori}, \citenamefont {Ruostekoski}, \citenamefont {Denz},\ and\ \citenamefont {Dennis}}]{sugic2021particle}%
  \BibitemOpen
  \bibfield  {author} {\bibinfo {author} {\bibfnamefont {D.}~\bibnamefont {Sugic}}, \bibinfo {author} {\bibfnamefont {R.}~\bibnamefont {Droop}}, \bibinfo {author} {\bibfnamefont {E.}~\bibnamefont {Otte}}, \bibinfo {author} {\bibfnamefont {D.}~\bibnamefont {Ehrmanntraut}}, \bibinfo {author} {\bibfnamefont {F.}~\bibnamefont {Nori}}, \bibinfo {author} {\bibfnamefont {J.}~\bibnamefont {Ruostekoski}}, \bibinfo {author} {\bibfnamefont {C.}~\bibnamefont {Denz}}, \ and\ \bibinfo {author} {\bibfnamefont {M.~R.}\ \bibnamefont {Dennis}},\ }\href@noop {} {\bibfield  {journal} {\bibinfo  {journal} {Nature Communications}\ }\textbf {\bibinfo {volume} {12}},\ \bibinfo {pages} {6785} (\bibinfo {year} {2021})}\BibitemShut {NoStop}%
\bibitem [{\citenamefont {Kuratsuji}\ and\ \citenamefont {Tsuchida}(2021)}]{kuratsuji2021evolution}%
  \BibitemOpen
  \bibfield  {author} {\bibinfo {author} {\bibfnamefont {H.}~\bibnamefont {Kuratsuji}}\ and\ \bibinfo {author} {\bibfnamefont {S.}~\bibnamefont {Tsuchida}},\ }\href@noop {} {\bibfield  {journal} {\bibinfo  {journal} {Physical Review A}\ }\textbf {\bibinfo {volume} {103}},\ \bibinfo {pages} {023514} (\bibinfo {year} {2021})}\BibitemShut {NoStop}%
\bibitem [{\citenamefont {Chen}\ \emph {et~al.}(2025)\citenamefont {Chen}, \citenamefont {Forbes},\ and\ \citenamefont {Qiu}}]{chen2025more}%
  \BibitemOpen
  \bibfield  {author} {\bibinfo {author} {\bibfnamefont {J.}~\bibnamefont {Chen}}, \bibinfo {author} {\bibfnamefont {A.}~\bibnamefont {Forbes}}, \ and\ \bibinfo {author} {\bibfnamefont {C.-W.}\ \bibnamefont {Qiu}},\ }\href@noop {} {\bibfield  {journal} {\bibinfo  {journal} {Light: Science \& Applications}\ }\textbf {\bibinfo {volume} {14}},\ \bibinfo {pages} {28} (\bibinfo {year} {2025})}\BibitemShut {NoStop}%
\bibitem [{\citenamefont {Wang}\ \emph {et~al.}(2024)\citenamefont {Wang}, \citenamefont {Zhao}, \citenamefont {Ma}, \citenamefont {Cai}, \citenamefont {Zhang}, \citenamefont {Shang}, \citenamefont {Zhang}, \citenamefont {Qin}, \citenamefont {Pong}, \citenamefont {Marozs{\'a}k} \emph {et~al.}}]{wang2024topological}%
  \BibitemOpen
  \bibfield  {author} {\bibinfo {author} {\bibfnamefont {A.~A.}\ \bibnamefont {Wang}}, \bibinfo {author} {\bibfnamefont {Z.}~\bibnamefont {Zhao}}, \bibinfo {author} {\bibfnamefont {Y.}~\bibnamefont {Ma}}, \bibinfo {author} {\bibfnamefont {Y.}~\bibnamefont {Cai}}, \bibinfo {author} {\bibfnamefont {R.}~\bibnamefont {Zhang}}, \bibinfo {author} {\bibfnamefont {X.}~\bibnamefont {Shang}}, \bibinfo {author} {\bibfnamefont {Y.}~\bibnamefont {Zhang}}, \bibinfo {author} {\bibfnamefont {J.}~\bibnamefont {Qin}}, \bibinfo {author} {\bibfnamefont {Z.-K.}\ \bibnamefont {Pong}}, \bibinfo {author} {\bibfnamefont {T.}~\bibnamefont {Marozs{\'a}k}},  \emph {et~al.},\ }\href@noop {} {\bibfield  {journal} {\bibinfo  {journal} {Light: Science \& Applications}\ }\textbf {\bibinfo {volume} {13}},\ \bibinfo {pages} {314} (\bibinfo {year} {2024})}\BibitemShut {NoStop}%
\bibitem [{\citenamefont {Ornelas}\ \emph {et~al.}(2025)\citenamefont {Ornelas}, \citenamefont {Nape}, \citenamefont {de~Mello~Koch},\ and\ \citenamefont {Forbes}}]{ornelas2025topological}%
  \BibitemOpen
  \bibfield  {author} {\bibinfo {author} {\bibfnamefont {P.}~\bibnamefont {Ornelas}}, \bibinfo {author} {\bibfnamefont {I.}~\bibnamefont {Nape}}, \bibinfo {author} {\bibfnamefont {R.}~\bibnamefont {de~Mello~Koch}}, \ and\ \bibinfo {author} {\bibfnamefont {A.}~\bibnamefont {Forbes}},\ }\href@noop {} {\bibfield  {journal} {\bibinfo  {journal} {Nature Communications}\ }\textbf {\bibinfo {volume} {16}},\ \bibinfo {pages} {2934} (\bibinfo {year} {2025})}\BibitemShut {NoStop}%
\bibitem [{\citenamefont {Shen}\ \emph {et~al.}(2024)\citenamefont {Shen}, \citenamefont {He}, \citenamefont {Song}, \citenamefont {Chen}, \citenamefont {He}, \citenamefont {Ma}, \citenamefont {Fells}, \citenamefont {Elston}, \citenamefont {Morris}, \citenamefont {Booth} \emph {et~al.}}]{shen2024topologically}%
  \BibitemOpen
  \bibfield  {author} {\bibinfo {author} {\bibfnamefont {Y.}~\bibnamefont {Shen}}, \bibinfo {author} {\bibfnamefont {C.}~\bibnamefont {He}}, \bibinfo {author} {\bibfnamefont {Z.}~\bibnamefont {Song}}, \bibinfo {author} {\bibfnamefont {B.}~\bibnamefont {Chen}}, \bibinfo {author} {\bibfnamefont {H.}~\bibnamefont {He}}, \bibinfo {author} {\bibfnamefont {Y.}~\bibnamefont {Ma}}, \bibinfo {author} {\bibfnamefont {J.~A.}\ \bibnamefont {Fells}}, \bibinfo {author} {\bibfnamefont {S.~J.}\ \bibnamefont {Elston}}, \bibinfo {author} {\bibfnamefont {S.~M.}\ \bibnamefont {Morris}}, \bibinfo {author} {\bibfnamefont {M.~J.}\ \bibnamefont {Booth}},  \emph {et~al.},\ }\href@noop {} {\bibfield  {journal} {\bibinfo  {journal} {Physical Review Applied}\ }\textbf {\bibinfo {volume} {21}},\ \bibinfo {pages} {024025} (\bibinfo {year} {2024})}\BibitemShut {NoStop}%
\bibitem [{\citenamefont {He}\ \emph {et~al.}(2024)\citenamefont {He}, \citenamefont {Meng}, \citenamefont {Wang}, \citenamefont {Zhong}, \citenamefont {Mata-Cervera}, \citenamefont {Li}, \citenamefont {Yan}, \citenamefont {Liu}, \citenamefont {Shen},\ and\ \citenamefont {Xiao}}]{he2024optical}%
  \BibitemOpen
  \bibfield  {author} {\bibinfo {author} {\bibfnamefont {T.}~\bibnamefont {He}}, \bibinfo {author} {\bibfnamefont {Y.}~\bibnamefont {Meng}}, \bibinfo {author} {\bibfnamefont {L.}~\bibnamefont {Wang}}, \bibinfo {author} {\bibfnamefont {H.}~\bibnamefont {Zhong}}, \bibinfo {author} {\bibfnamefont {N.}~\bibnamefont {Mata-Cervera}}, \bibinfo {author} {\bibfnamefont {D.}~\bibnamefont {Li}}, \bibinfo {author} {\bibfnamefont {P.}~\bibnamefont {Yan}}, \bibinfo {author} {\bibfnamefont {Q.}~\bibnamefont {Liu}}, \bibinfo {author} {\bibfnamefont {Y.}~\bibnamefont {Shen}}, \ and\ \bibinfo {author} {\bibfnamefont {Q.}~\bibnamefont {Xiao}},\ }\href@noop {} {\bibfield  {journal} {\bibinfo  {journal} {Nature Communications}\ }\textbf {\bibinfo {volume} {15}},\ \bibinfo {pages} {10141} (\bibinfo {year} {2024})}\BibitemShut {NoStop}%
\bibitem [{\citenamefont {Lin}\ \emph {et~al.}(2024{\natexlab{a}})\citenamefont {Lin}, \citenamefont {Ota}, \citenamefont {Arakawa},\ and\ \citenamefont {Iwamoto}}]{lin2024chip}%
  \BibitemOpen
  \bibfield  {author} {\bibinfo {author} {\bibfnamefont {W.}~\bibnamefont {Lin}}, \bibinfo {author} {\bibfnamefont {Y.}~\bibnamefont {Ota}}, \bibinfo {author} {\bibfnamefont {Y.}~\bibnamefont {Arakawa}}, \ and\ \bibinfo {author} {\bibfnamefont {S.}~\bibnamefont {Iwamoto}},\ }\href@noop {} {\bibfield  {journal} {\bibinfo  {journal} {Optica}\ }\textbf {\bibinfo {volume} {11}},\ \bibinfo {pages} {1588} (\bibinfo {year} {2024}{\natexlab{a}})}\BibitemShut {NoStop}%
\bibitem [{\citenamefont {Shen}\ \emph {et~al.}(2022)\citenamefont {Shen}, \citenamefont {Mart{\'\i}nez},\ and\ \citenamefont {Rosales-Guzm{\'a}n}}]{shen2022generation}%
  \BibitemOpen
  \bibfield  {author} {\bibinfo {author} {\bibfnamefont {Y.}~\bibnamefont {Shen}}, \bibinfo {author} {\bibfnamefont {E.~C.}\ \bibnamefont {Mart{\'\i}nez}}, \ and\ \bibinfo {author} {\bibfnamefont {C.}~\bibnamefont {Rosales-Guzm{\'a}n}},\ }\href@noop {} {\bibfield  {journal} {\bibinfo  {journal} {ACS Photonics}\ }\textbf {\bibinfo {volume} {9}},\ \bibinfo {pages} {296} (\bibinfo {year} {2022})}\BibitemShut {NoStop}%
\bibitem [{\citenamefont {Mitra}\ \emph {et~al.}(2025)\citenamefont {Mitra}, \citenamefont {Madasu}, \citenamefont {Gabardos}, \citenamefont {Kwong}, \citenamefont {Shen}, \citenamefont {Ruostekoski},\ and\ \citenamefont {Wilkowski}}]{mitra2025topological}%
  \BibitemOpen
  \bibfield  {author} {\bibinfo {author} {\bibfnamefont {C.}~\bibnamefont {Mitra}}, \bibinfo {author} {\bibfnamefont {C.~S.}\ \bibnamefont {Madasu}}, \bibinfo {author} {\bibfnamefont {L.}~\bibnamefont {Gabardos}}, \bibinfo {author} {\bibfnamefont {C.~C.}\ \bibnamefont {Kwong}}, \bibinfo {author} {\bibfnamefont {Y.}~\bibnamefont {Shen}}, \bibinfo {author} {\bibfnamefont {J.}~\bibnamefont {Ruostekoski}}, \ and\ \bibinfo {author} {\bibfnamefont {D.}~\bibnamefont {Wilkowski}},\ }\href@noop {} {\bibfield  {journal} {\bibinfo  {journal} {APL Photonics}\ }\textbf {\bibinfo {volume} {10}} (\bibinfo {year} {2025})}\BibitemShut {NoStop}%
\bibitem [{\citenamefont {Lin}\ \emph {et~al.}(2024{\natexlab{b}})\citenamefont {Lin}, \citenamefont {Liu}, \citenamefont {Duan}, \citenamefont {Du},\ and\ \citenamefont {Yuan}}]{lin2024wavelength}%
  \BibitemOpen
  \bibfield  {author} {\bibinfo {author} {\bibfnamefont {M.}~\bibnamefont {Lin}}, \bibinfo {author} {\bibfnamefont {Q.}~\bibnamefont {Liu}}, \bibinfo {author} {\bibfnamefont {H.}~\bibnamefont {Duan}}, \bibinfo {author} {\bibfnamefont {L.}~\bibnamefont {Du}}, \ and\ \bibinfo {author} {\bibfnamefont {X.}~\bibnamefont {Yuan}},\ }\href@noop {} {\bibfield  {journal} {\bibinfo  {journal} {Applied Physics Reviews}\ }\textbf {\bibinfo {volume} {11}} (\bibinfo {year} {2024}{\natexlab{b}})}\BibitemShut {NoStop}%
\bibitem [{\citenamefont {Liu}\ \emph {et~al.}()\citenamefont {Liu}, \citenamefont {Ma}, \citenamefont {Yang}, \citenamefont {Liu}, \citenamefont {Chen}, \citenamefont {Li}, \citenamefont {Song}, \citenamefont {Qiu}, \citenamefont {Zou}, \citenamefont {Hu} \emph {et~al.}}]{liu2025nanophotonic}%
  \BibitemOpen
  \bibfield  {author} {\bibinfo {author} {\bibfnamefont {J.}~\bibnamefont {Liu}}, \bibinfo {author} {\bibfnamefont {J.}~\bibnamefont {Ma}}, \bibinfo {author} {\bibfnamefont {J.}~\bibnamefont {Yang}}, \bibinfo {author} {\bibfnamefont {S.}~\bibnamefont {Liu}}, \bibinfo {author} {\bibfnamefont {B.}~\bibnamefont {Chen}}, \bibinfo {author} {\bibfnamefont {X.}~\bibnamefont {Li}}, \bibinfo {author} {\bibfnamefont {C.}~\bibnamefont {Song}}, \bibinfo {author} {\bibfnamefont {G.}~\bibnamefont {Qiu}}, \bibinfo {author} {\bibfnamefont {K.}~\bibnamefont {Zou}}, \bibinfo {author} {\bibfnamefont {X.}~\bibnamefont {Hu}},  \emph {et~al.},\ }\href {\doibase 10.1038/s41567-025-02973-y} {\bibfield  {journal} {\bibinfo  {journal} {Nature Physics}\ }10.1038/s41567-025-02973-y}\BibitemShut {NoStop}%
\bibitem [{\citenamefont {Ornelas}\ \emph {et~al.}(2024)\citenamefont {Ornelas}, \citenamefont {Nape}, \citenamefont {de~Mello~Koch},\ and\ \citenamefont {Forbes}}]{ornelas2024non}%
  \BibitemOpen
  \bibfield  {author} {\bibinfo {author} {\bibfnamefont {P.}~\bibnamefont {Ornelas}}, \bibinfo {author} {\bibfnamefont {I.}~\bibnamefont {Nape}}, \bibinfo {author} {\bibfnamefont {R.}~\bibnamefont {de~Mello~Koch}}, \ and\ \bibinfo {author} {\bibfnamefont {A.}~\bibnamefont {Forbes}},\ }\href@noop {} {\bibfield  {journal} {\bibinfo  {journal} {Nature Photonics}\ }\textbf {\bibinfo {volume} {18}},\ \bibinfo {pages} {258} (\bibinfo {year} {2024})}\BibitemShut {NoStop}%
\bibitem [{Sup()}]{Supplement}%
  \BibitemOpen
  \href@noop {} {}\bibinfo {note} {Supplementary Material, \href{https://doi.org/10.1103/xm7z-bnpl}{https://doi.org/10.1103/xm7z-bnpl}, contains details about the spin-orbit device, state derivation, experimental methods, and GHZ proposal, which includes Refs.~ \cite{brasselet_prl_2018,leach2009violation,agnew2011tomography}}\BibitemShut {NoStop}%
\bibitem [{\citenamefont {Yan}\ \emph {et~al.}(2015)\citenamefont {Yan}, \citenamefont {Gregg}, \citenamefont {Karimi}, \citenamefont {Rubano}, \citenamefont {Marrucci}, \citenamefont {Boyd},\ and\ \citenamefont {Ramachandran}}]{yan2015q}%
  \BibitemOpen
  \bibfield  {author} {\bibinfo {author} {\bibfnamefont {L.}~\bibnamefont {Yan}}, \bibinfo {author} {\bibfnamefont {P.}~\bibnamefont {Gregg}}, \bibinfo {author} {\bibfnamefont {E.}~\bibnamefont {Karimi}}, \bibinfo {author} {\bibfnamefont {A.}~\bibnamefont {Rubano}}, \bibinfo {author} {\bibfnamefont {L.}~\bibnamefont {Marrucci}}, \bibinfo {author} {\bibfnamefont {R.}~\bibnamefont {Boyd}}, \ and\ \bibinfo {author} {\bibfnamefont {S.}~\bibnamefont {Ramachandran}},\ }\href@noop {} {\bibfield  {journal} {\bibinfo  {journal} {Optica}\ }\textbf {\bibinfo {volume} {2}},\ \bibinfo {pages} {900} (\bibinfo {year} {2015})}\BibitemShut {NoStop}%
\bibitem [{\citenamefont {Brasselet}(2018)}]{brasselet_prl_2018}%
  \BibitemOpen
  \bibfield  {author} {\bibinfo {author} {\bibfnamefont {E.}~\bibnamefont {Brasselet}},\ }\href@noop {} {\bibfield  {journal} {\bibinfo  {journal} {Physical Review Letters}\ }\textbf {\bibinfo {volume} {121}},\ \bibinfo {pages} {033901} (\bibinfo {year} {2018})}\BibitemShut {NoStop}%
\bibitem [{\citenamefont {Verstraete}\ and\ \citenamefont {Wolf}(2002)}]{verstraete2002entanglement}%
  \BibitemOpen
  \bibfield  {author} {\bibinfo {author} {\bibfnamefont {F.}~\bibnamefont {Verstraete}}\ and\ \bibinfo {author} {\bibfnamefont {M.~M.}\ \bibnamefont {Wolf}},\ }\href@noop {} {\bibfield  {journal} {\bibinfo  {journal} {Physical review letters}\ }\textbf {\bibinfo {volume} {89}},\ \bibinfo {pages} {170401} (\bibinfo {year} {2002})}\BibitemShut {NoStop}%
\bibitem [{\citenamefont {Toninelli}\ \emph {et~al.}(2019)\citenamefont {Toninelli}, \citenamefont {Ndagano}, \citenamefont {Vall{\'e}s}, \citenamefont {Sephton}, \citenamefont {Nape}, \citenamefont {Ambrosio}, \citenamefont {Capasso}, \citenamefont {Padgett},\ and\ \citenamefont {Forbes}}]{toninelli2019concepts}%
  \BibitemOpen
  \bibfield  {author} {\bibinfo {author} {\bibfnamefont {E.}~\bibnamefont {Toninelli}}, \bibinfo {author} {\bibfnamefont {B.}~\bibnamefont {Ndagano}}, \bibinfo {author} {\bibfnamefont {A.}~\bibnamefont {Vall{\'e}s}}, \bibinfo {author} {\bibfnamefont {B.}~\bibnamefont {Sephton}}, \bibinfo {author} {\bibfnamefont {I.}~\bibnamefont {Nape}}, \bibinfo {author} {\bibfnamefont {A.}~\bibnamefont {Ambrosio}}, \bibinfo {author} {\bibfnamefont {F.}~\bibnamefont {Capasso}}, \bibinfo {author} {\bibfnamefont {M.~J.}\ \bibnamefont {Padgett}}, \ and\ \bibinfo {author} {\bibfnamefont {A.}~\bibnamefont {Forbes}},\ }\href@noop {} {\bibfield  {journal} {\bibinfo  {journal} {Advances in Optics and Photonics}\ }\textbf {\bibinfo {volume} {11}},\ \bibinfo {pages} {67} (\bibinfo {year} {2019})}\BibitemShut {NoStop}%
\bibitem [{\citenamefont {Nagali}\ \emph {et~al.}(2009)\citenamefont {Nagali}, \citenamefont {Sciarrino}, \citenamefont {De~Martini}, \citenamefont {Marrucci}, \citenamefont {Piccirillo}, \citenamefont {Karimi},\ and\ \citenamefont {Santamato}}]{nagali2009quantum}%
  \BibitemOpen
  \bibfield  {author} {\bibinfo {author} {\bibfnamefont {E.}~\bibnamefont {Nagali}}, \bibinfo {author} {\bibfnamefont {F.}~\bibnamefont {Sciarrino}}, \bibinfo {author} {\bibfnamefont {F.}~\bibnamefont {De~Martini}}, \bibinfo {author} {\bibfnamefont {L.}~\bibnamefont {Marrucci}}, \bibinfo {author} {\bibfnamefont {B.}~\bibnamefont {Piccirillo}}, \bibinfo {author} {\bibfnamefont {E.}~\bibnamefont {Karimi}}, \ and\ \bibinfo {author} {\bibfnamefont {E.}~\bibnamefont {Santamato}},\ }\href@noop {} {\bibfield  {journal} {\bibinfo  {journal} {Physical Review Letters}\ }\textbf {\bibinfo {volume} {103}},\ \bibinfo {pages} {013601} (\bibinfo {year} {2009})}\BibitemShut {NoStop}%
\bibitem [{\citenamefont {Brasselet}(2021)}]{brasselet_bookchapter_2021}%
  \BibitemOpen
  \bibfield  {author} {\bibinfo {author} {\bibfnamefont {E.}~\bibnamefont {Brasselet}},\ }\href@noop {} {\bibfield  {journal} {\bibinfo  {journal} {Liquid Crystals: New Perspectives (eds Pieranski P. and Godinho MH).(Hoboken: John Wiley \& Sons, 2021)}\ ,\ \bibinfo {pages} {1}} (\bibinfo {year} {2021})}\BibitemShut {NoStop}%
\bibitem [{\citenamefont {Leach}\ \emph {et~al.}(2009)\citenamefont {Leach}, \citenamefont {Jack}, \citenamefont {Romero}, \citenamefont {Ritsch-Marte}, \citenamefont {Boyd}, \citenamefont {Jha}, \citenamefont {Barnett}, \citenamefont {Franke-Arnold},\ and\ \citenamefont {Padgett}}]{leach2009violation}%
  \BibitemOpen
  \bibfield  {author} {\bibinfo {author} {\bibfnamefont {J.}~\bibnamefont {Leach}}, \bibinfo {author} {\bibfnamefont {B.}~\bibnamefont {Jack}}, \bibinfo {author} {\bibfnamefont {J.}~\bibnamefont {Romero}}, \bibinfo {author} {\bibfnamefont {M.}~\bibnamefont {Ritsch-Marte}}, \bibinfo {author} {\bibfnamefont {R.}~\bibnamefont {Boyd}}, \bibinfo {author} {\bibfnamefont {A.}~\bibnamefont {Jha}}, \bibinfo {author} {\bibfnamefont {S.}~\bibnamefont {Barnett}}, \bibinfo {author} {\bibfnamefont {S.}~\bibnamefont {Franke-Arnold}}, \ and\ \bibinfo {author} {\bibfnamefont {M.}~\bibnamefont {Padgett}},\ }\href@noop {} {\bibfield  {journal} {\bibinfo  {journal} {Optics express}\ }\textbf {\bibinfo {volume} {17}},\ \bibinfo {pages} {8287} (\bibinfo {year} {2009})}\BibitemShut {NoStop}%
\bibitem [{\citenamefont {Agnew}\ \emph {et~al.}(2011)\citenamefont {Agnew}, \citenamefont {Leach}, \citenamefont {McLaren}, \citenamefont {Roux},\ and\ \citenamefont {Boyd}}]{agnew2011tomography}%
  \BibitemOpen
  \bibfield  {author} {\bibinfo {author} {\bibfnamefont {M.}~\bibnamefont {Agnew}}, \bibinfo {author} {\bibfnamefont {J.}~\bibnamefont {Leach}}, \bibinfo {author} {\bibfnamefont {M.}~\bibnamefont {McLaren}}, \bibinfo {author} {\bibfnamefont {F.~S.}\ \bibnamefont {Roux}}, \ and\ \bibinfo {author} {\bibfnamefont {R.~W.}\ \bibnamefont {Boyd}},\ }\href@noop {} {\bibfield  {journal} {\bibinfo  {journal} {Physical Review A—Atomic, Molecular, and Optical Physics}\ }\textbf {\bibinfo {volume} {84}},\ \bibinfo {pages} {062101} (\bibinfo {year} {2011})}\BibitemShut {NoStop}%
\end{thebibliography}
\end{document}